\documentclass[pre,twocolumn,preprintnumbers,superscriptaddress,nofootinbib]{revtex4-1} 
\usepackage{blindtext}
\usepackage{amsmath}
\usepackage{amsfonts}
\usepackage{graphicx}
\usepackage{float}
\usepackage{booktabs}
\usepackage{gensymb}
\usepackage{hyperref}
\usepackage[caption=false]{subfig}
\usepackage{ragged2e}
\DeclareCaptionJustification{justified}{\justifying}
\usepackage{courier}
\usepackage{soul}
\usepackage{subfig}
\usepackage{times}
\usepackage{amsthm}
\usepackage{amssymb}
\usepackage{listings}
\usepackage{color}

\usepackage[usestackEOL]{stackengine}
\usepackage{scalerel}
\usepackage{graphicx,amsmath}
\parskip \baselineskip
\def\myupbracefill#1{\rotatebox{90}{\stretchto{\{}{#1}}}
\def\rlwd{.5pt}
\newcommand\notate[4][B]{%
  \if B#1\else\def\myupbracefill##1{}\fi%
  \def\useanchorwidth{T}%
  \setbox0=\hbox{$\displaystyle#2$}%
  \def\stackalignment{c}\stackunder[-6pt]{%
    \def\stackalignment{c}\stackunder[-1.5pt]{%
      \stackunder[2pt]{\strut $\displaystyle#2$}{\myupbracefill{\wd0}}}{%
    \rule{\rlwd}{#3\baselineskip}}}{%
  \strut\kern9pt$\rightarrow$\smash{\rlap{$~\displaystyle#4$}}}%
}

\definecolor{dkgreen}{rgb}{0,0.6,0}
\definecolor{gray}{rgb}{0.5,0.5,0.5}
\definecolor{mauve}{rgb}{0.58,0,0.82}

\definecolor{mygreen}{RGB}{28,172,0} 
\definecolor{mylilas}{RGB}{170,55,241}

\usepackage{bm}
\usepackage{fancyhdr}
\usepackage{slashed}
\usepackage{latexsym,epsfig,bbm}

\newcommand{\bra}[1]{\langle{#1}|}
\newcommand{\ket}[1]{|{#1}\rangle}

\newcommand{\braket}[2]{\langle{#1}|{#2}\rangle}
\newcommand{\bopk}[3]{\langle{#1}|{#2}|{#3}\rangle}

\newcommand{\figref}[1]{Fig.~\ref{#1}}

\definecolor{blue}{rgb}{0,0.2,1}

\definecolor{red}{rgb}{0.9,0,0}

\newcommand{\past}[1]{\overleftarrow{#1}}
\newcommand{\fut}[1]{\overrightarrow{#1}}
\newcommand{\pastfut}[1]{\overleftrightarrow{#1}}

\begin{document}

\title{Robust inference of memory structure for efficient quantum modelling of stochastic processes}

\author{Matthew Ho}
\email{hosh0021@e.ntu.edu.sg}
\affiliation{School of Physical and Mathematical Sciences, Nanyang Technological University, Singapore 637371, Singapore}
\affiliation{Complexity Institute, Nanyang Technological University, Singapore 637335, Singapore}

\author{Mile Gu}
\email{mgu@quantumcomplexity.org}
\affiliation{School of Physical and Mathematical Sciences, Nanyang Technological University, Singapore 637371, Singapore}
\affiliation{Complexity Institute, Nanyang Technological University, Singapore 637335, Singapore}
\affiliation{Centre for Quantum Technologies. National University of Singapore, 3 Science Drive 2, Singapore 117543, Singapore}

\author{Thomas J. Elliott}
\email{physics@tjelliott.net}
\affiliation{Complexity Institute, Nanyang Technological University, Singapore 637335, Singapore}
\affiliation{School of Physical and Mathematical Sciences, Nanyang Technological University, Singapore 637371, Singapore}

\date{\today}

\begin{abstract}

A growing body of work has established the modelling of stochastic processes as a promising area of application for quantum techologies; it has been shown that quantum models are able to replicate the future statistics of a stochastic process whilst retaining less information about the past than any classical model must -- even for a purely classical process. Such memory-efficient models open a potential future route to study complex systems in greater detail than ever before, and suggest profound consequences for our notions of structure in their dynamics. Yet, to date methods for constructing these quantum models are based on having a prior knowledge of the optimal classical model. Here, we introduce a protocol for blind inference of the memory structure of quantum models  -- tailored to take advantage of quantum features --  direct from time-series data, in the process highlighting the robustness of their structure to noise. This in turn provides a way to construct memory-efficient quantum models of stochastic processes whilst circumventing certain drawbacks that manifest solely as a result of classical information processing in classical inference protocols.

\end{abstract}

\maketitle

\section{Introduction}

Complex processes are prevalent throughout the world, taking the form of natural processes such as the weather~\cite{Lynch2008, Bauer2015} and DNA sequences~\cite{Pavlos2015}, as well as artificial processes like the stock market~\cite{Preis2012} and traffic~\cite{Kerner1996}. We construct models of these processes in order to better understand their structure and predict their behaviour. Within complexity science, the field of computational mechanics~\cite{Crutchfield1989, shalizi2001computational, Crutchfield2011} offers a systematic approach to understanding the intrinsic computation of a process by identifying the causal links between its past and future, and has been used to study a diverse set of dynamics such as deterministic chaos in the logistic map~\cite{Crutchfield1989, Crutchfield2011}, cellular automata~\cite{Hanson1997}, the dripping faucet experiment~\cite{Goncalves1998}, stock markets~\cite{Park2007}, and neural spike trains~\cite{Haslinger2010}. A key component of the approach are so-called $\varepsilon$-\emph{machines}, which as a valuable byproduct represent the most parsimonius causal model of a process.

In recent decades, the prospect of using quantum effects in information processing has emerged, promising advantages for a range of applications in terms of algorithmic speed-ups~\cite{montanaro2016quantum}, secure communication~\cite{bennett2014quantum}, and beyond. Stochastic modelling is no exception to this, and a growing body of work has established that when information is encoded into a quantum memory, causal models of a stochastic process can be designed that function whilst retaining less information about the past than is classically possible~\cite{Gu2012, Mahoney2016, Riechers2016, Thompson2017, Elliott2018a, Binder2018, Elliott2018b, Liu2019}. This quantum memory advantage can grow unbounded~\cite{Garner2017a, Aghamohammadi2017, Elliott2018a, Elliott2018b, Thompson2018a, elliott2019extreme}, and has been verified experimentally~\cite{Palsson2017a, Jouneghani2017, Ghafari2018}. Like its classical counterpart, the amount of information stored within these quantum models has been suggested as a measure of structural complexity in stochastic dynamics~\cite{Tan2014, Suen2017, Aghamohammadi2017a, Suen2018}.

\begin{figure}

	\includegraphics[width=\linewidth]{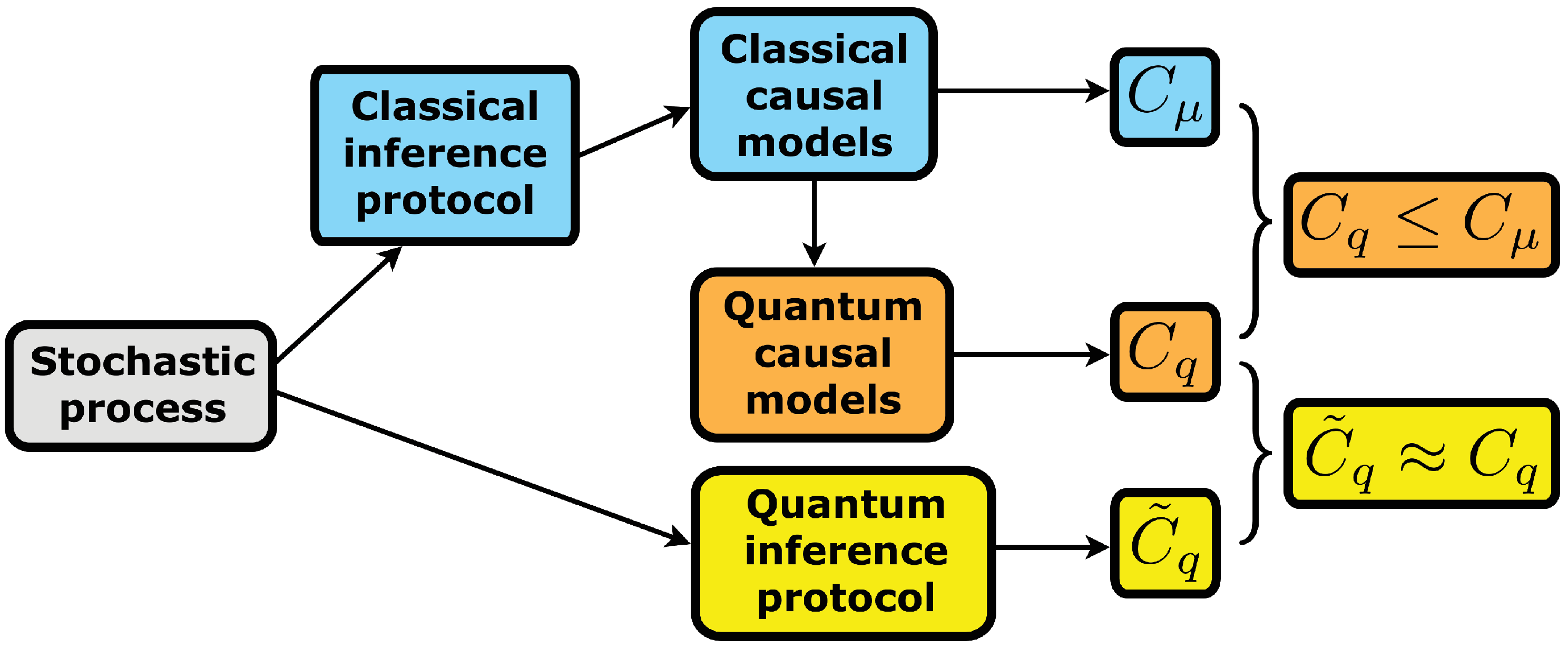}
	\caption{{\bf Schematic context of our work.} Quantum information processing has been shown to provide a more memory efficient route to stochastic modelling than classically possible. However, current approaches to constructing quantum models first require classical models to be inferred; here we introduce a blind inference protocol for going straight from raw data to quantum structure. The quantities $C_\mu$, $C_q$ and $\tilde{C}_q$ represent the information stored by the minimal classical, quantum, and inferred quantum causal models respectively.}
	\label{fig.framework}
\end{figure}

Currently, systematic approaches to constructing such quantum models are predicated on having a prior exact statistical description of the process, or knowledge of its $\varepsilon$-machine. As a result, to apply these tools to real-world systems we must first use classical inference protocols to construct an $\varepsilon$-machine~\cite{Crutchfield1989, Shalizi2004, Strelioff2014}, and then use this as a basis to construct a corresponding quantum model. It is desirable to instead have a model inference protocol to directly go from data to the quantum model, avoiding any extra computational overhead associated with also determining the classical model. In this vein, here we introduce such a protocol for directly inferring the memory structure of a quantum model of a stochastic process -- which we show is robust to statistical noise. The protocol is tailored specifically for quantum models, taking advantage of certain of their features that allow some approximations that must be made in classical information processing to be avoided. \figref{fig.framework} provides a schematic of our motivation.

The layout of this article is as follows. In Section \ref{sec.framework} we outline the general framework of stochastic processes and computational mechanics as is relevant here, as well as the more efficient quantum models. Section \ref{ref.proofs} provides the core of our results, introducing the inference protocol, showing its robustness to statistical fluctuations, and justifying its accuracy. The efficacy of our inference protocol is then demonstrated in practice with two toy processes in Section \ref{sec.examples}. Finally, we conclude in Section \ref{sec.discussion}, and discuss some future directions.

\section{Framework}
\label{sec.framework}

\subsection{Stochastic processes}
\label{sec.processes}

We consider discrete-time stochastic processes represented by a bi-infinite probabalistic string of outcomes $\overleftrightarrow{X} \equiv X_{-\infty:\infty} = \ldots X_{-2} X_{-1} X_0 X_1 X_2\ldots$, where $X_t$ are random variables that take on values $x_t$ drawn from an alphabet $\mathcal{A}$, and the subscript $t$ represents the timestep. Consecutive strings $X_{0:t} := X_{0}X_{1}\ldots X_{t-1}$ are called \textit{words}, with the left index inclusive and the right exclusive. We consider stationary processes, such that  $P(X_{0:L}) = P(X_{t:t+L}) \ \forall \ t, L \in \mathbb{Z}$. We partition the process into (semi-infinite) pasts and futures, denoted as  $\overleftarrow{x} \equiv x_{-\infty:0}$ and $\overrightarrow{x} \equiv x_{0:\infty}$ respectively, where $t=0$ is taken to be the present.

The \emph{Markov order} is an important property of a process that defines an effective history length; a process is said to have Markov order $R$ if $R$ is the smallest value such that $P(X_0|\past{X}) =  P(X_0|X_{-R:0})$ is satisfied~\cite{Racca2007}. That is, it is the smallest block length of the most recent past that provides a sufficient statistic of the future. When $R=1$ the process is said to be Markovian.


\subsection{Models}
\label{sec.computationalmechanics}

\textbf{Computational mechanics.} Computational mechanics~\cite{Crutchfield1989, shalizi2001computational, Crutchfield2011} provides a formal statistical framework for identifying and analysing structure in complex processes. We outline key elements of computational mechanics here, providing further detail in Appendix A. Its modus operandi involves the minimal causal representation\footnote{Here, by causal we mean that the representation stores no information about the future of the process that could not be deduced from past observations.} of stochastic processes, which may be determined by a systematic clustering of pasts. Specifically, the \emph{causal states} of a process are a set of equivalence classes on the pasts, defined according to the relation
\begin{equation}
\label{eq.equivalencerelation}
	\overleftarrow{x} \sim_\varepsilon \overleftarrow{x}' \Leftrightarrow P(\overrightarrow{X}|\overleftarrow{X} = \overleftarrow{x}) = P(\overrightarrow{X}|\overleftarrow{X} = \overleftarrow{x}');
\end{equation}
that is, two pasts belong to the same causal state iff they give rise to statistically-identical futures. We label the causal states as $s_j \in \mathcal{S}$.

Because of the deterministic assignment of pasts to causal states, it can be seen that transitions between causal states are also deterministic conditional on the output symbol; that is, given a past $\past{x}\in s_j$, upon emission of the next symbol $x$ the new past $\past{x}x$ must belong to causal state $s_{\lambda(x,j)}$, where $\lambda(x,j)$ is a deterministic update function. This deterministic transition structure is sometimes referred to as \emph{unifilarity}, and allows us to represent the process as a deterministic edge-emitting hidden Markov model (HMM) known as the $\varepsilon$-machine, where the causal states form the hidden states of the model, and the edge emissions the observed symbols~\cite{Crutchfield1989, shalizi2001computational}.

The amount of information stored by the $\varepsilon$-machine can be quantified by the Shannon entropy of the stationary distribution on causal states:
\begin{equation}
\label{eq.Cmu}
	C_\mu :=H[P(s_j)]= - \sum_{s_j\in\mathcal{S}} P(s_j) \log_2 [P(s_j)],
\end{equation}
where $P(s_j)=\sum_{\past{x}\in s_j}P(\past{x})$. Across all (classical) causal representations of a process, the $\varepsilon$-machine minimises the information cost of its corresponding memory states, and it is in this sense we refer to it as being minimal (or optimal). Because of this distinguished feature, $C_\mu$ is called the \emph{statistical complexity}, and is considered as a quantifier of structure in the process~\cite{Crutchfield2011}, in some sense representing how much information about the past is needed to produce the future. This quantity is lower bounded by the mutual information between the past and future $I(\overleftarrow{X};\overrightarrow{X})$\cite{shalizi2001computational}; in general this bound is not strict, and the difference is referred to as the modelling overhead~\cite{Mahoney2011a}.

When dealing with raw data, one must estimate the probabilities through inference, which will be subject to unavoidable statistical fluctuations due to the finite amount of data. As such, when applying the equivalence relation Eq.~\eqref{eq.equivalencerelation} a threshold $\delta$-tolerence must be permitted, where pasts are assigned to the same causal state if their conditional future distributions are `close enough'~\cite{Crutchfield1989}\footnote{It should be noted that there is not a fixed definition of how this tolerence should be implemented, but typically it would be appropriate to use some form of statistical distance between the conditional distributions, with some maximal allowed distance for merging parameterised by $\delta$.}. Adjusting the strictness of this tolerence induces diferent levels of coarse-graining: if too narrow then the fluctuations will lead to additional spurious causal states that would have been merged with knowledge of the exact distributions; if too loose then pasts with different conditional futures can be merged. Depending on this degree of coarse-graining, the obtained value of $C_\mu$ will vary; the statistical complexity is sensitive to statistical fluctuations, and is generally not robust to noise.

\begin{figure}

	\includegraphics[width=\linewidth]{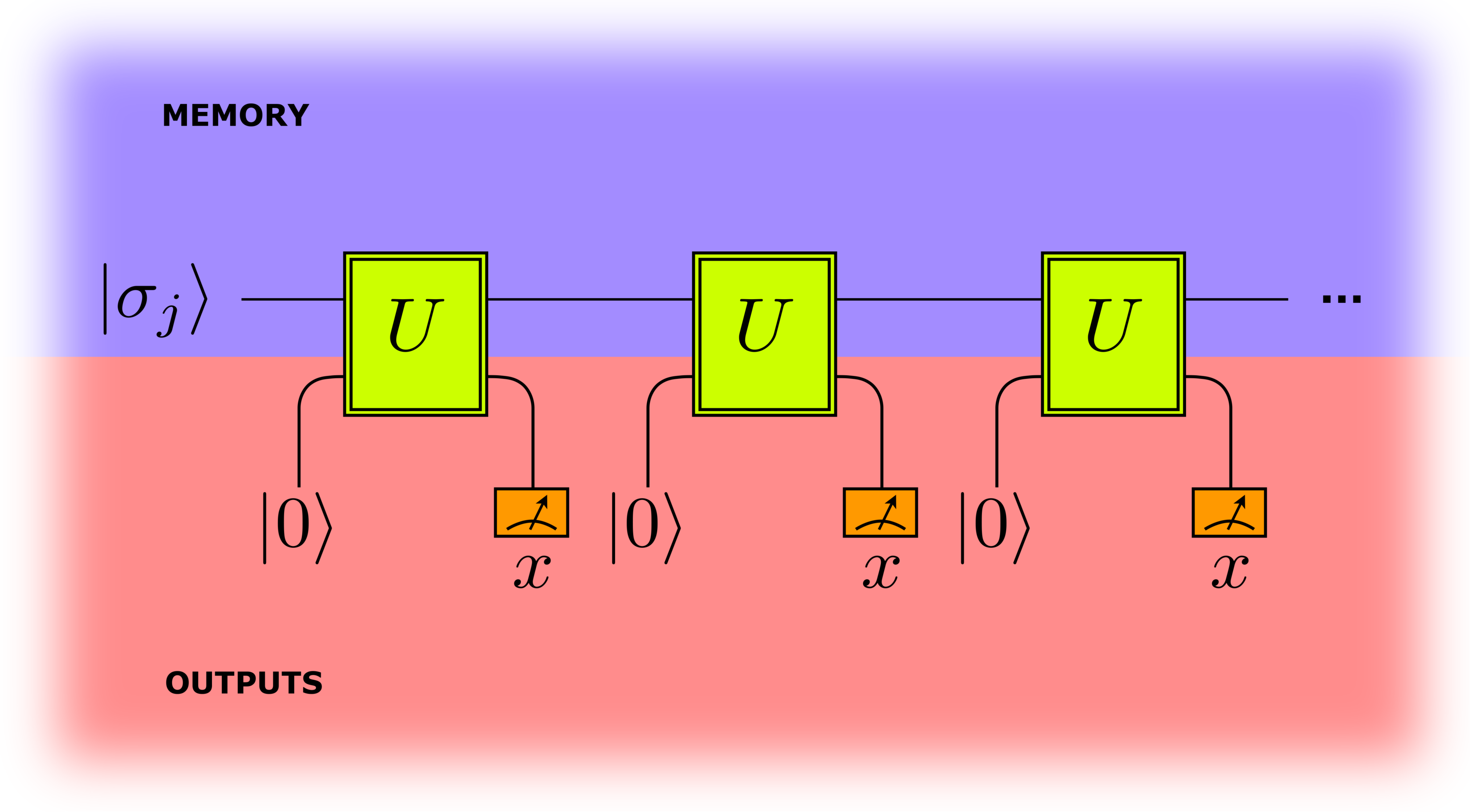}
	\caption{\textbf{Unitary quantum models of stochastic processes.} Repeated unitary interactions between a quantum memory and probe ancilla produces a string of stochastic outputs when the probe is measured. The specific form of the interaction depends on the particular process being modelled; the output statistics will then be as specified by this process.}
	\label{Fig.unitaryoperation}
\end{figure}

\textbf{Quantum computational mechanics.} When considering quantum methods of information processing, the minimality of $\varepsilon$-machines no longer holds; it has been shown that causal quantum models can be found with lower information costs~\cite{Gu2012}, using non-orthogonal memory states to reduce the modelling overhead. 
The current state-of-the-art quantum models~\cite{Liu2019} are based on unitary interactions between the memory subsystem and a probe ancilla:
\begin{equation}
\label{eq.QMS}
	U \ket{\sigma_j} \ket{0} = \sum_{x \in \mathcal{A}} \sqrt{P(x|s_j)} e^{i\varphi_{xj}} \ket{\sigma_{\lambda(x,j)}} \ket{x},
\end{equation}
where the first subspace contains the memory, the second the probe (measured after the interaction to produce the symbol for that timestep), $\{\ket{\sigma_j}\}$ are the quantum memory states (in one-to-one correspondence with the causal states $\{s_j\}$), and the phase factors $\{\varphi_{xj}\}$ are tunable parameters. Successive applications of $U$ on the quantum memory state and probe at each timestep will yield a string of outputs from the probe measurement that are statistically distributed according to the modelled process\footnote{After each timestep, the probe ancilla is either reset, or a fresh ancilla is introduced.}, as depicted in \figref{Fig.unitaryoperation}.

The corresponding memory cost is called the \emph{quantum statistical memory}, given by the von Neumann entropy of the quantum memory states:
\begin{equation}
\label{eq.Cq}
	C_q :=S[\rho] = -\mathrm{tr}(\rho \log_2 [\rho]),
\end{equation}
where $\rho=\sum_{s_j\in\mathcal{S}}P(s_j)\ket{\sigma_j}\bra{\sigma_j}$. The title of quantum statistical complexity is reserved for the minimum of this quantity over all causal quantum models; there is as yet no systematic approach to finding this minimal model however, and optimising over the phase factors is a cumbersome task. For this reason, we shall here use the best quantum models for which a systematic construction method is known: \emph{phaseless} unitary quantum models~\cite{Binder2018}, given by Eq.~\eqref{eq.QMS} with all phase factors set to zero. Despite not generally being minimal, the corresponding quantum statistical memory of these models has still been suggested as quantifier of structure~\cite{Gu2012, Tan2014, Suen2017, Aghamohammadi2017a, Suen2018}, often emphasising different features to the classical statistical complexity.


\section{Inference protocol}
\label{ref.proofs}

We here introduce an inference protocol for the quantum statistical memory $\tilde{C}_q$\footnote{We use tildes to represent estimated quantities.} of the phaseless unitary quantum models that can be used to investigate structure in time-series data. The protocol is tailored specifically to take advantage of features of the specific model, and bypassing the need to construct the $\varepsilon$-machine as an intermediate step. It is agnostic to the causal architecture of the process, and requires only an estimate of the Markov order.

The inference protocol is based on a set of postulated quantum memory states given by clustering pasts in which the last $L$ symbols are identical; we thus have a memory state $\ket{\varsigma_{x_{-L:0}}}$ for each of the possible $L$-length words $x_{-L:0}$. The choice of $L$ should correspond to the estimated Markov order of the process (or at least, the effective Markov order -- see below). From the data we then estimate the conditional probabilities $\tilde{P}(X_0|X_{-L:0})$, and implicitly define the quantum memory states to satisfy the interaction
\begin{equation} \label{eq.inferredQMS}
	U \ket{\varsigma_{x_{-L:0}}} \ket{0} = \sum_{x_0 \in \mathcal{A}} \sqrt{\tilde{P}(x_{0}|x_{-L:0})} \ket{\varsigma_{x_{-L+1:1}}} \ket{x_0}.
\end{equation}
A set of quantum memory states and corresponding interaction can be found through a recursive expression for the overlaps of the states and employing a reverse Gram-Schmidt procedure~\cite{Binder2018}. The estimated quantum statistical memory is then given by the von Neumann entropy of the corresponding stationary state of the memory:
\begin{equation}
	\tilde{C}_q := S[\rho^{(L)}]= -\text{tr}(\rho^{(L)} \log_2 [\rho^{(L)}]),
\end{equation}
where $\rho^{(L)} = \sum_{x_{-L:0}} \tilde{P}(x_{-L:0}) \ket{\varsigma_{x_{-L:0}}} \bra{\varsigma_{x_{-L:0}}}$.

To show that this protocol provides a faithful estimate of the quantum statistical memory, we first prove two properties of the above construction:
\begin{description}
	\item[A] \emph{Self-merging of quantum memory states}\\We show that when $L$ is at least as large as the Markov order the overlap of quantum memory states assigned to different pasts in the same causal state is unity when exact probabilities are used.
	\item[B] \emph{Robustness of quantum statistical memory.}\\We show that the quantum statistical memory of the phaseless unitary quantum model is insensitive to small perturbations in the probabilities.
\end{description}
With these properties we can then discuss the accuracy of the inference protocol:
\begin{description}
	\item[C] \emph{Inference protocol.}\\We indicate how the accuracy of our estimate of quantum statistical memory scales with the amount of data, and how it converges for sufficiently large data streams.
\end{description}

\subsection{Self-merging of quantum memory states}
\label{sec.selfmerging}

We first show that the blind construction Eq.~\eqref{eq.inferredQMS} will automatically adopt the causal architecture of the process (i.e., that pasts belonging to the same causal state are assigned to the same memory state) without explicit need to apply the causal equivalence relation~\cite{Elliott2018b}, provided that exact probabilities are used, and the chosen $L$ is at least as large as the Markov order of the process. That is, the quantum memory states we construct will correspond to the same states as would be obtained from the phaseless form of the model Eq.~\eqref{eq.QMS}, but without prior knowledge of how the pasts are clustered into causal states. In turn, this means the blind construction will faithfully replicate the process, with the same quantum statistical memory.

To see this, let $R$ denote the Markov order of the process, and recall that this means $P(\fut{X}|X_{-R:0})=P(\fut{X}|\past{X})$. Since the Markov order can alternatively be expressed as the longest history length needed to determine the causal state (i.e., $R = \mathrm{min}\{ r: H(S_0|x_{-r:0}) = 0\})$, all pasts where the latest $R$ symbols are identical belong to the same causal state~\cite{Mahoney2016}. We can see that if $L\geq R$, the construction already correctly merges all pasts where the latest $L$ symbols are identical.

By analogy with the corresponding methods for phaseless unitary models~\cite{Binder2018}, we can express
\begin{align}
\label{eq.overlap}
\braket{\varsigma_{x_{-L:0}}}{\varsigma_{x_{-L:0}'}}&=\bopk{\varsigma_{x_{-L:0}}}{U^\dagger U}{\varsigma_{x_{-L:0}'}}\nonumber\\
&\hspace{-3.5em}=\!\!\! \sum_{x_0 \in \mathcal{A}}\!\! \sqrt{P(x_{0}|x_{-L:0})P(x_{0}|x_{-L:0}')} \braket{\varsigma_{x_{-L+1:1}}}{\varsigma_{x_{-L+1:1}'}}.
\end{align}
Iteratively applying this relation, we obtain that
\begin{equation}
\label{eq.overlap2}
\braket{\varsigma_{x_{-L:0}}}{\varsigma_{x_{-L:0}'}}=\sum_{\fut{x}}\!\! \sqrt{P(\fut{x}|x_{-L:0})P(\fut{x}|x_{-L:0}')};
\end{equation}
if $L\geq R$ we then have
\begin{equation}
\label{eq.overlap3}
\braket{\varsigma_{x_{-L:0}}}{\varsigma_{x_{-L:0}'}}=\sum_{\fut{x}}\!\! \sqrt{P(\fut{x}|\past{x})P(\fut{x}|\past{x}')},
\end{equation}
where the full pasts $\past{x}$ and $\past{x}'$ can be taken as any pasts with the correct corresponding last $L$ symbols. We are thus able to conclude that 
\begin{equation}
\label{eq.QMSequiv}
\braket{\varsigma_{x_{-L:0}}}{\varsigma_{x_{-L:0}'}}=1\Leftrightarrow P(\fut{X}|\past{x})=P(\fut{X}|\past{x}'),
\end{equation}
which can be seen as an instantiation of the causal equivalence relation, i.e., two pasts are mapped to the same memory state iff they have the same conditional future statistics.

\subsection{Robustness of quantum statistical memory}
\label{sec.robustness}

We next show that the quantum statistical memory of our construction is robust to small perturbations of the probabilities. Consider mapping  
\begin{equation}
\label{eq.perturb}
P(\pastfut{X})\to P^\epsilon(\pastfut{X})=P(\pastfut{X})+\epsilon\Delta P(\pastfut{X}),
\end{equation}
where $\Delta P$ governs the relative changes in the distribution for each string, and $\epsilon$ the strength of the perturbation. We here outline a proof that the perturbation to $C_q$ scales smoothly with $\epsilon$; full details may be found in Appendix B.

The Gram matrix $G$ of a quantum state $\rho=\sum_j P_j\ket{\sigma_j}\bra{\sigma_j}$ is defined as $G_{jk}=\sqrt{P_jP_k}\braket{\sigma_j}{\sigma_k}$, and can be shown to have the same spectrum as $\rho$~\cite{Jozsa2000, Horn2012, Riechers2016}. As such, it is possible to define the Gram matrix of our construction as
\begin{equation}
\label{eq.gram}
	G_{x_{-L:0}x'_{-L:0}} = \sqrt{P(x_{-L:0}) P(x_{-L:0}')} \braket{\varsigma_{x_{-L:0}}}{\varsigma_{x'_{-L:0}}},
\end{equation}
and correspondingly, from its spectrum calculate $C_q$.

Consider that we have $L\geq R$, such that it is possible to express the overlaps of the quantum memory states as 
\begin{equation}
\label{eq.overlap4}
\braket{\varsigma_{x_{-L:0}}}{\varsigma_{x_{-L:0}'}}=\sum_{x_{0:L}}\!\! \sqrt{P(x_{0:L}|x_{-L:0})P(x_{0:L}|x_{-L:0}')}.
\end{equation}
Note that we only need consider $L$ steps into the future as this uniquely determines the subsequent memory state, independent of the past. Now replace each of the probability distributions in this expression by their corresponding perturbed forms, which may be obtained from the marginals of Eq.~\eqref{eq.perturb}. We can then calculate the perturbed form of the corresponding Gram matrix using this expression for the overlaps, and show that its spectrum varies smoothly with $\epsilon$. Since the von Neumann entropy is a continuous function of the spectrum of a state~\cite{Nielsen2010}, we thus find that it too smoothly deforms with $\epsilon$. 

Hence, we can conclude that the quantum statistical memory is robust to small perturbations in the probability distributions. Due to the self-merging of our quantum memory states, we can also see that the quantum statistical memory of phaseless unitary quantum models is similarly robust in general. This is in contrast to the classical statistical complexity, which can vary discontinuously with the probabilities -- notably, whenever the pertubation triggers a new merging of pasts into a causal state, or conversely, the splitting of a causal state.

\subsection{Inference protocol}
\label{sec.inferenceprotocol}

With these two results in hand, we are now in a position to argue the accuracy of our inference protocol. By parsing the data and adopting a frequentist approach to estimate the marginal distribution of words of length $L+1$ (for some chosen $L$), we are able to estimate the conditional probabilities needed for Eq.~\eqref{eq.inferredQMS}. Moreover, we can use these marginals to construct estimates for the probabilities needed to evaluate Eq.~\eqref{eq.gram} (the conditional probabilities $\tilde{P}(X_{0:L}|X_{-L:0})$ can be obtained from multiplying $\tilde{P}(X_0|X_{-L:0})$ with the assumption that $L$ is at least as large of the Markov order), and in turn, estimate $C_q$. From the results of the previous two subsections, we can be assured that this should be an faithful estimate, provided that the estimated marginals are close to the exact distributions, and $L\geq R$.

Let us first consider the error in our estimates of the marginals. With any inference from finite data there will of course be statistical fluctuations; this is not problematic for our inference protocol provided that these fluctuations are sufficiently small. These fluctuations can be treated in the same manner as the perturbations of the previous section, but with the perturbation terms $\Delta P$ now being stochastic variables.  In Appendix C we show that it the size of the error in the inferred Gram matrix and the estimated quantum statistical memory $\tilde{C}_q$ approximately scale as $O(|\mathcal{A}|^L/\sqrt{N})$, where $|\mathcal{A}|$ is the size of the alphabet, and $N$ the size of the data stream.  

The next question is how we determine the choice of $L$. From the above, we see that taking too large an $L$ will lead to untenably large fluctuations. We must therefore cap it at some value $L_{\mathrm{max}}$ where $|\mathcal{A}|^L_{\mathrm{max}}/\sqrt{N}\ll1$; as a rough guideline we suggest $L_{\mathrm{max}}\lesssim \log_{|\mathcal{A}|}(N/1,000)$. On the other hand, we require $L$ to be large enough that it matches or exceeds the Markov order of the process, in order to effect the (approximate) self-merging of quantum memory states.

When a process has a large, or even infinite Markov order, it may not be possible to have an $L$ that satisfies both of these requirements. Nevertheless, while the Markov order tells us how far back into the past memory effects can persist, it does not inform us how strong they are. It is often the case that these long-range historical dependencies only have minor influence on the future, and that the recent past is much more relevant. In such instances, the influence of the distant past can be thought of as a small perturbation to the statistics with respect to the recent past alone, and so from the previous subsection we can expect that they have minimal impact on the requisite quantum statistical memory. We therefore introduce the concept of an \emph{effective Markov order} $R_{\mathrm{eff}}$, that encapsulates the idea that a sufficiently-long string of past observations that is less than the Markov order may nevertheless still be `good enough' to capture most of the predictive information contained in the past.

We define the effective Markov order as the smallest length of a string of past observations for which the influence of considering an additional symbol one step further into the past does not exceed some threshold. Specifically, we define $R_{\mathrm{eff}}$ as the smallest integer $r$ that satisfies
\begin{equation}
\label{eq.emo}
\max_{xx'} \langle D(P(X_0|xX_{-r:0}),P(X_0|x'X_{-r:0})\rangle<\delta,
\end{equation}
where $\delta$ is the parameter defining the threshold\footnote{Strictly, we have a family of effective Markov orders for the process, parameterised by $\delta$.}, the expectation value is taken over the distribution of strings of length $r$, and $D$ is some distance measure between probability distributions; for the purposes of this work we will use the trace distance $D(P,Q):=\sum_x |P(x)-Q(x)|/2$. We can estimate the effective Markov order from this data, and use this as a guide for choosing a value for $L$ in the inference protocol.


\section{Examples}
\label{sec.examples}

We now demonstrate the efficacy of our model with two toy example processes: the so-called $R$-$k$ golden mean and nemo processes. We use an exact HMM representation of the processes to generate a representative string of outputs, and infer $\tilde{C}_q$ directly from this time-series. We choose the initial state according to the stationary distribution of the HMM, such that the output statistics are representative of the stationary state of the process. We will look at how the estimate for the quantum statistical memory varies both with the amount of data $N$, and the history length $L$, highlighting the range of $L$ values that would be considered appropriate given the above considerations regarding the (effective) Markov order and expected size of fluctuations.

\subsection{Golden mean process}

\begin{figure}
\includegraphics[width=\linewidth]{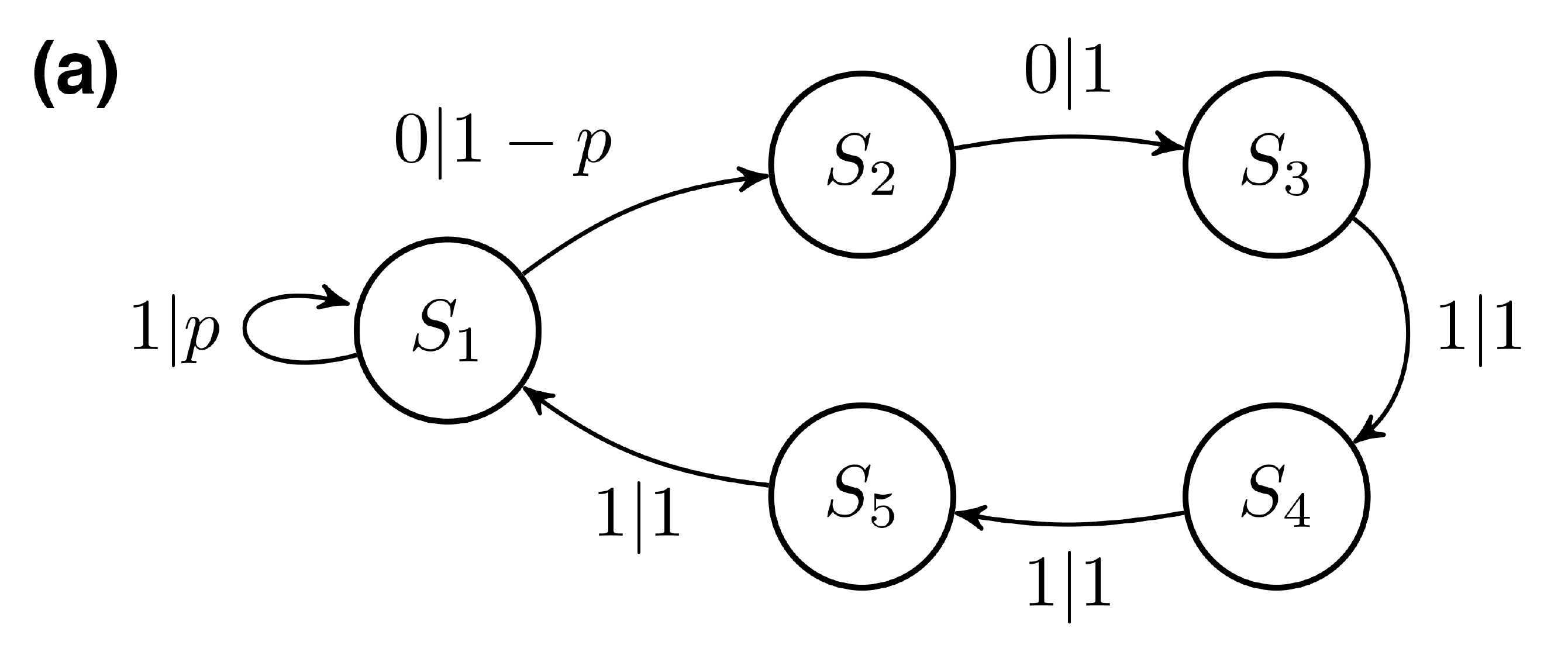}		
\includegraphics[width=\linewidth]{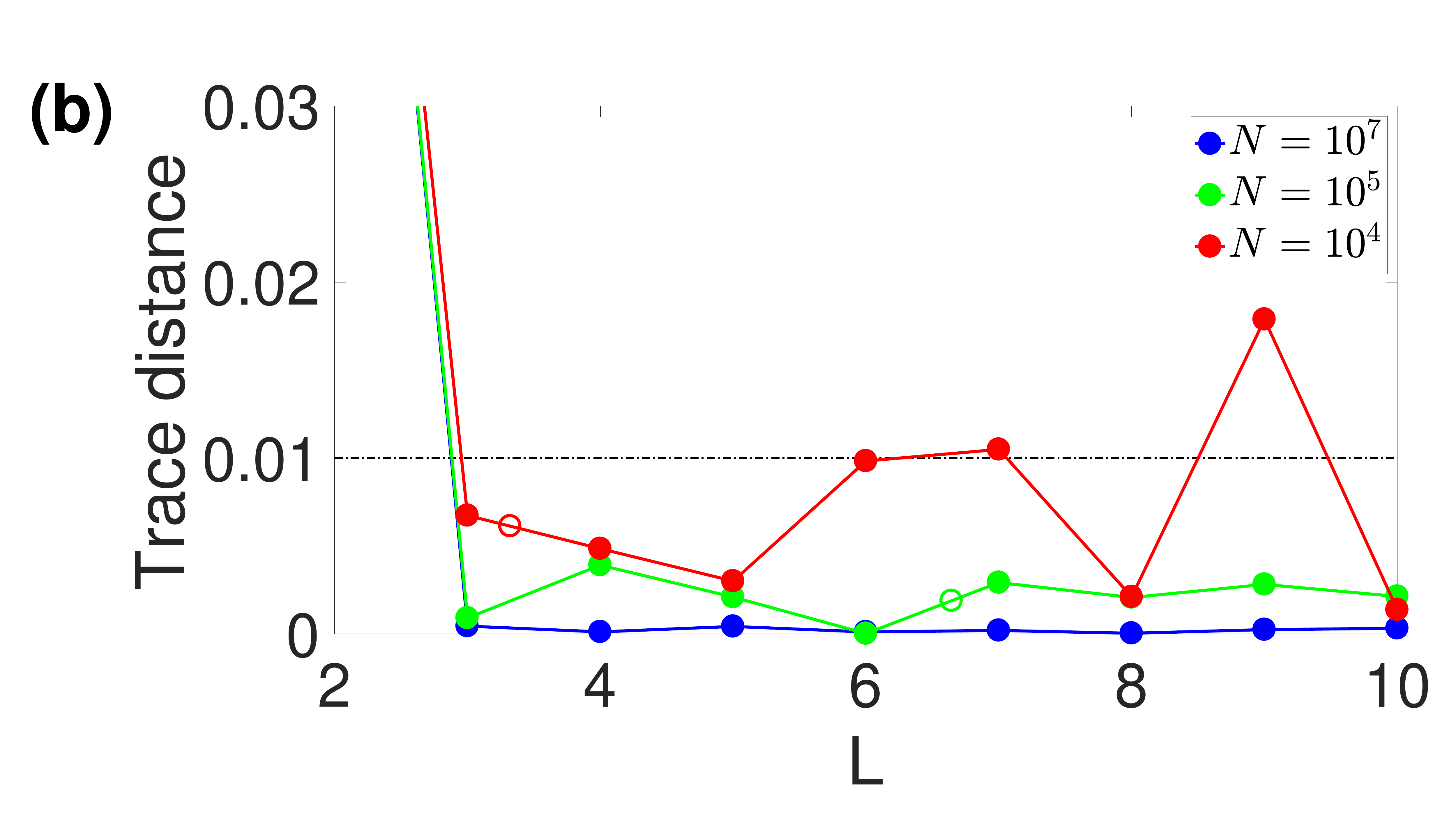}
\includegraphics[width=\linewidth]{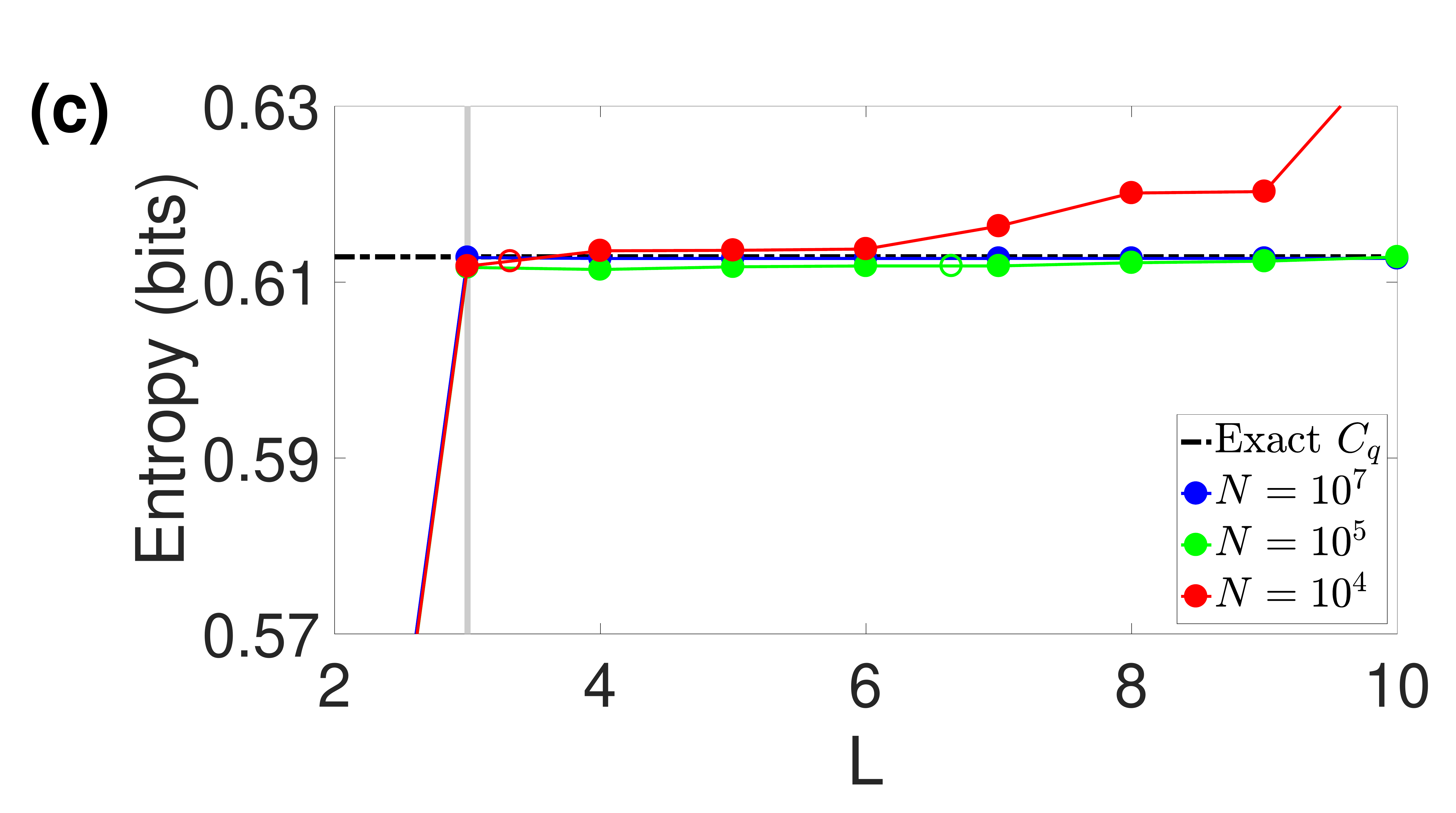}
\caption{{\bf Golden mean process.} (a) $\varepsilon$-machine for the 3-2 golden mean process; the notation $x|P$ denotes that the indicated transition between states involves output of symbol $x$ and occurs with probability $P$. (b) Average trace distance for variation $L+1$ steps in the past; hollow circles denote the points where $L=\log_2(N/1,000)$, and the dashed line $\delta=0.01$. (c) Comparison of exact and estimated quantum statistical memory for different lengths of data stream; the variation with $L$ is shown for the estimated quantities, and the vertical line indicates the effective Markov order for $\delta=0.01$. For plots (b) and (c) we take $p=0.9$.}
\label{GoldenMeanProcess}
\end{figure}

We first look at the $R$-$k$ golden mean process family. Here, $R$ and $k$ are tunable parameters that correspond to the Markov order and cryptic order\footnote{The cryptic order is a counterpart to the Markov order, describing the minimum length of past observations required to be certain of the present causal state, given that one knows the entire future, i.e., the smallest $k$ satisfying $H[S|X_{-k:\infty}]=0$.} of the process respectively~\cite{Crutchfield2009a, Mahoney2009, Mahoney2011a, Mahoney2016}. Here, we will consider the $3$-$2$ golden mean process specifically, as represented by its $\varepsilon$-machine in \figref{GoldenMeanProcess}(a).

In \figref{GoldenMeanProcess}(b) we plot the expectation of the trace distance between differing symbols increasingly far into the past (i.e, between $\tilde{P}(X_0|0X_{-R:0})$ and $\tilde{P}(X_0|1X_{-R:0}$), from which we can infer an effective Markov order for the process as defined in Eq.~\eqref{eq.emo}. The limits of finite data are already visible in this plot, with the instability clear when $L$ is too large relative to $N$, due to undersampling of the process statistics. The hollow circle on each plot represents the point at which $L=\log_2(N/1,000)$ -- beyond this point we consider the statistics to be undersampled. Setting a threshold $\delta=0.01$, we would assign an effective Markov order of $R_{\mathrm{eff}}=3$, which aligns with the true Markov order of the process. \figref{GoldenMeanProcess}(c) displays the estimated $\tilde{C}_q$; we see that at the Markov order the estimate is very close to the exact value $C_q$, with statistical noise gradually degrading the quality of the estimate at larger $L$ when we have insufficient data. This highlights both the efficacy of protocol, and importance of selecting an appropriate value for $L$. The corresponding statistical complexity of the process $C_\mu \approx 1.435$ is omitted as it is much larger than both $C_q$ and $\tilde{C}_q$.

\subsection{Nemo process}

\begin{figure}
\includegraphics[width=\linewidth]{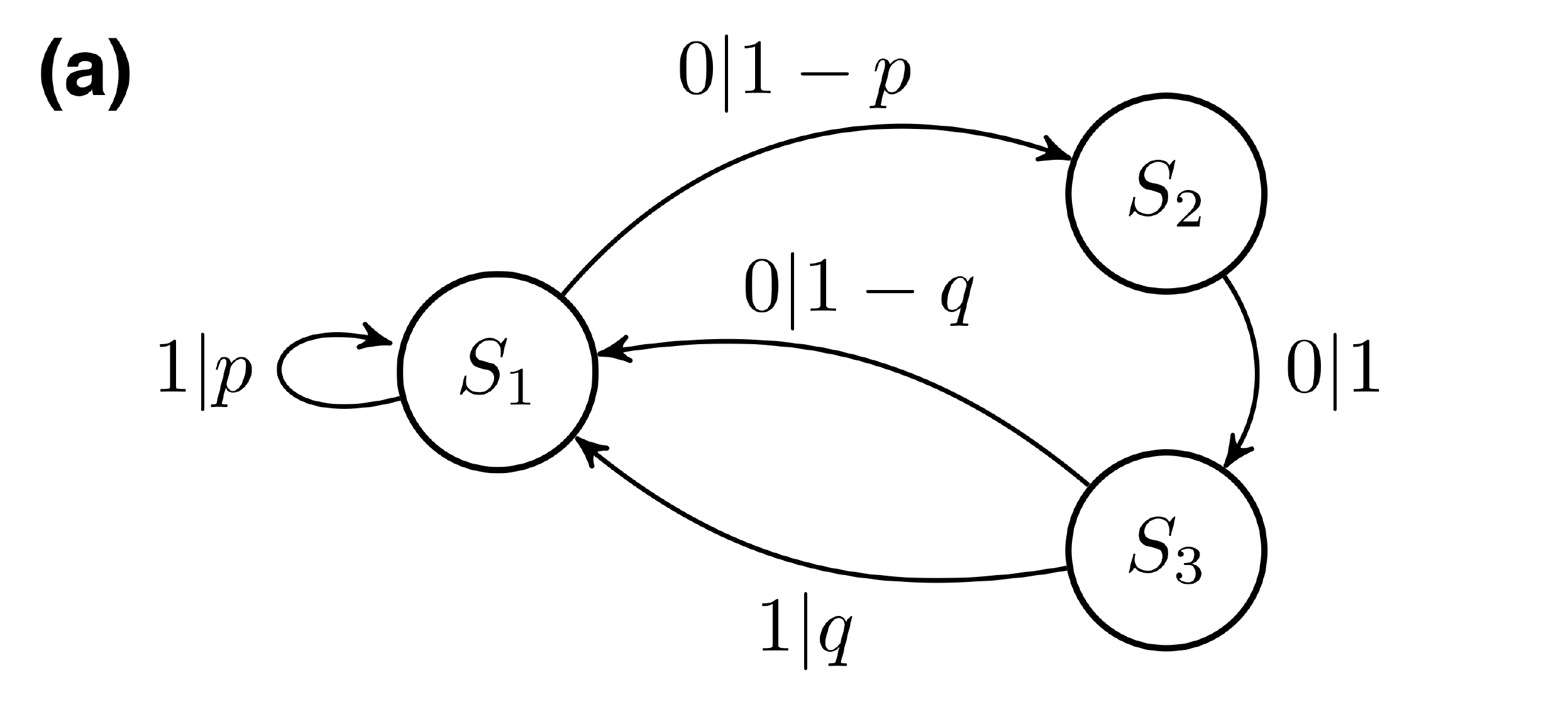}		
\includegraphics[width=\linewidth]{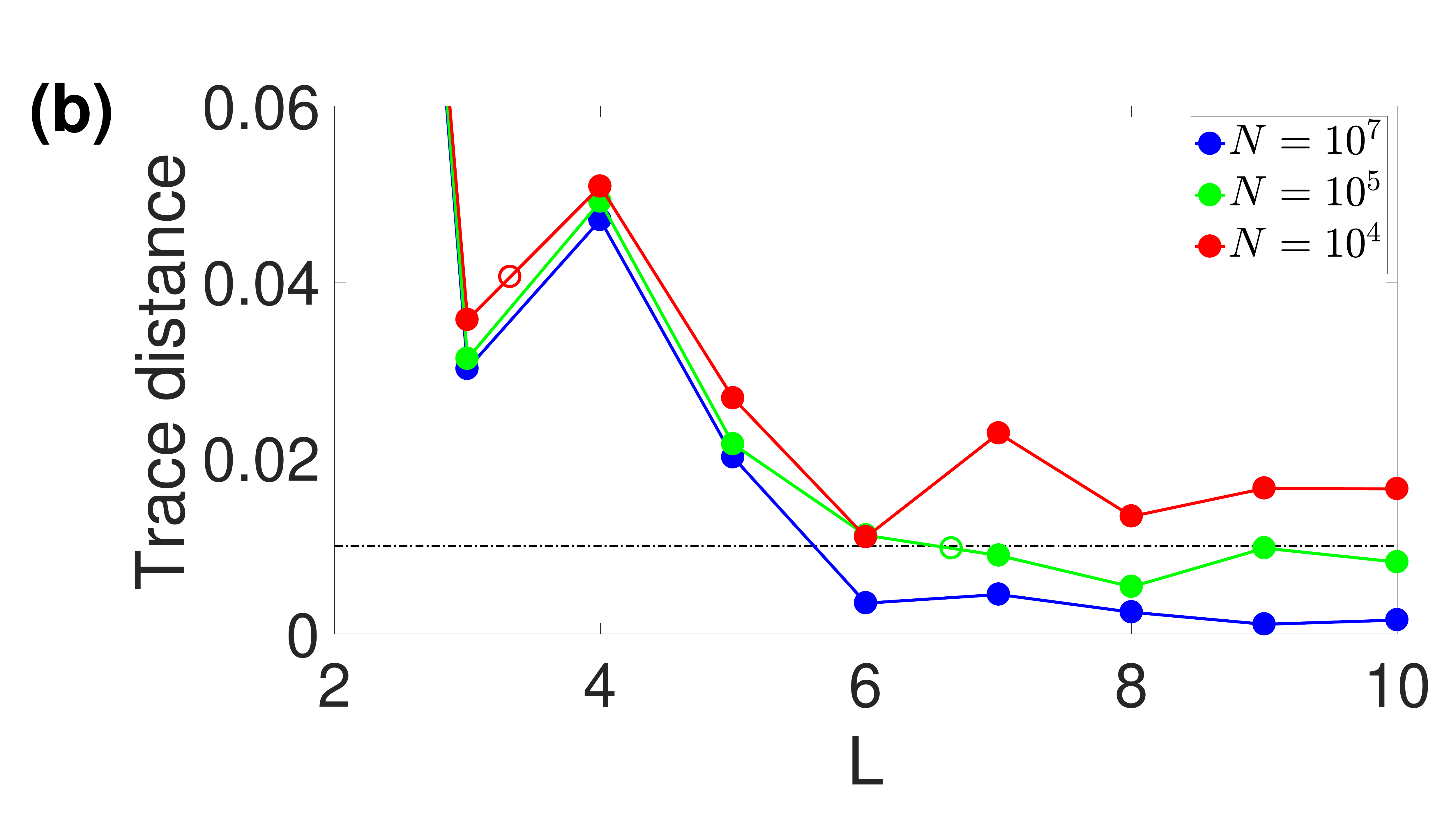}
\includegraphics[width=\linewidth]{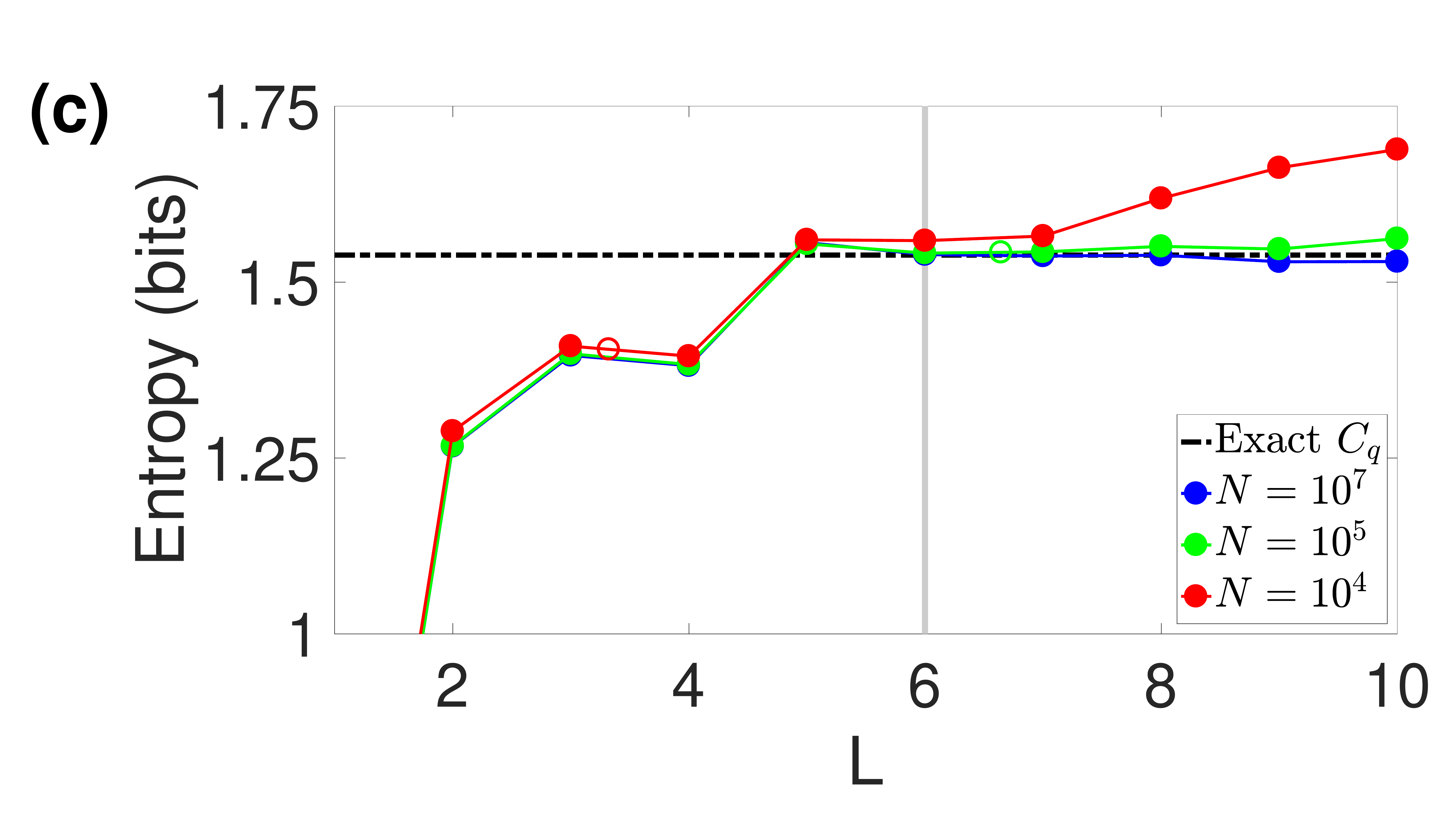}
\caption{{\bf Nemo process.} (a) $\varepsilon$-machine for the nemo process. (b) Average trace distance for variation $L+1$ steps in the past; hollow circles denote the points where $L=\log_2(N/1,000)$, and the dashed line $\delta=0.01$. (c) Comparison of exact and estimated quantum statistical memory for different lengths of data stream; the variation with $L$ is shown for the estimated quantities, and the vertical line indicates the effective Markov order for $\delta=0.01$. For plots (b) and (c) we take $p=0.1$ and $q=0.9$.}
	\label{NemoProcess}
\end{figure}

As a second example, we consider the nemo process, which can be represented by its $\varepsilon$-machine as in \figref{NemoProcess}(a). A key feature of this process is that it has infinite Markov order: a contiguous string of zeros of any length cannot be exactly synchronised to a causal state. As such, with this example it is not possible to choose an $L$ that matches the Markov order of the process. Nevertheless, we will show that the effective Markov order can provide a suitable proxy.

\figref{NemoProcess}(b) shows the expected trace distance for variation in the symbol $L+1$ steps into the past; setting a tolerance $\delta=0.01$ we assign an effective Markov order $R_{\mathrm{eff}}=6$. Examining the estimated quantum statistical memory $\tilde{C}_q$ [\figref{NemoProcess}(c)], we see that setting $L=6$ does indeed appear to provide an accurate estimate of $C_q$, striking a balance between allowing sufficiently long histories to capture most of the past dependency, while not going as far as to underfit. Note that in this case we should consider $N=10,000$ to be insufficient data to provide a good estimate, as the prescribed $L_{\mathrm{max}}$ is significantly smaller than the effective Markov order -- practically, this could be deduced from the trace distance, which never drops below the threshold value. The corresponding statistical complexity of the process $C_\mu \approx 1.583$ is again omitted from the figure.

\section{Discussion}
\label{sec.discussion}

We have introduced a protocol for estimating the information cost of quantum simulation of stochastic processes. We have shown that both this quantity and our protocol are robust to small statistical perturbations. This provides a means to characterise structure in a process according to the amount of (quantum) resources needed to capture its behaviour, analogous to corresponding classical quantities~\cite{Crutchfield1989, Shalizi2004, Strelioff2014}. Moreover, this provides a key step towards blind construction of quantum models that efficiently replicate the behaviour of such processes.

An essential consideration to be made in this latter vein is the capabilities of current and near-term quantum technologies. Our inferrence protocol accurately captures the information that must be stored by a quantum model of a process -- appropriately indicating the amount of structure in the process -- at the expense of indicating a multitude of memory states, typically $\sim|\mathcal{A}|^L$, in excess of the number of causal states. The number of memory states is parameterised by a companion metric, the \emph{topological memory} $D_q = \log_2[\text{dim}(\rho)]$; quantum advantages can also be found in terms of this measure~\cite{Thompson2018a, Ghafari2018, Liu2019, elliott2019extreme, loomis2019strong}. Future work will investigate methods of compression in this parameter via truncation in terms of the quantum state space, opening up the possibility to implement the inferred constructions experimentally. The accuracy of these inferred models can then be explored using recently-developed quantifiers of process distinguishability~\cite{yang2019measures}.

\acknowledgements
We thank Chew Lock Yue, Suen Whei Yeap, and Suryadi for discussions. This work was funded by the Lee Kuan Yew Endowment Fund (Postdoctoral Fellowship), Singapore Ministry of Education Tier 1 grant RG190/17, FQXi-RFP-1809 from the Foundational Questions Institute and Fetzer Franklin Fund, a donor advised fund of Silicon Valley Community Foundation, Singapore National Research Foundation Fellowship NRF-NRFF2016-02, and NRF-ANR grant NRF2017-NRF-ANR004 VanQuTe. T.J.E.~thanks the Centre for Quantum Technologies for their hospitality.

\widetext
\appendix

\section{Computational mechanics primer}

\begin{figure}[H]
\centering \includegraphics[width=0.25\linewidth]{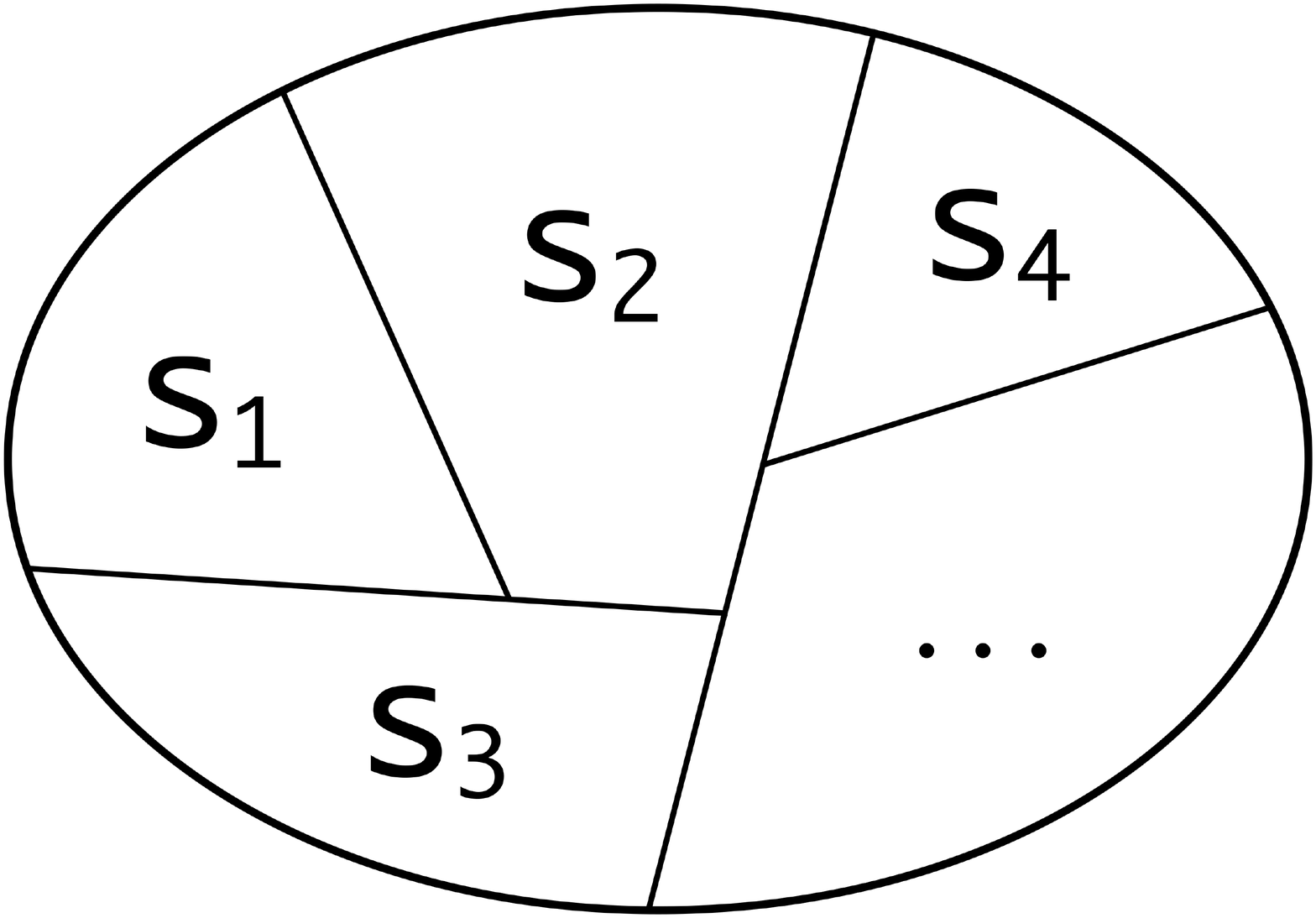}
	\caption{Pasts are clustered into groups -- causal states -- based on the causal equivalence relation Eq.~\eqref{eq.appequivalencerelation}; this illustration is sometimes termed `Occam's Pool'~\cite{shalizi2001computational}.}
	\label{fig.occamspool}
\end{figure}

Computational mechanics~\cite{Crutchfield1989, shalizi2001computational, Crutchfield2011} is a branch of complexity theory originating from studies of structure and intrinsic computation in dynamical systems and stochastic processes. The core of computational mechanics is in isolating information contained in the past of the process that is relevant to its future. This is achieved through the \emph{causal equivalence relation}:
\begin{equation}
\label{eq.appequivalencerelation}
	\overleftarrow{x} \sim_\varepsilon \overleftarrow{x}' \Leftrightarrow P(\overrightarrow{X}|\overleftarrow{X}=\overleftarrow{x}) = P(\overrightarrow{X}|\overleftarrow{X}=\overleftarrow{x}').
\end{equation}
This relation groups together two pasts $\overleftarrow{x}$ and $\overleftarrow{x}'$ iff their future outputs are statistically indistinguishable. The resulting equivalence classes are called \emph{causal states} $s_j\in\mathcal{S}$; this grouping is illustrated in \figref{fig.occamspool}. Put simply, if two pasts give rise to identical future statistics, retaining any further information that would distinguish between them yields no predictive power. It transpires that this is the optimal grouping of pasts: causal states are the minimal sufficient statistic of the past with respect to the future~\cite{shalizi2001computational}.

When considering bi-infinite, discrete-time stationary stochastic processes (as done in this work), the causal states can be represented as the latent states of an edge-emitting hidden Markov model (HMM), where the emissions on edges correspond to the output symbols of the process, and the probabilistic transition structure is defined by the process. Specifically, if a past $\past{x}\in s_j$ has a probability $P(x|j)$ of emitting $x$, and $\past{x}x\in s_k$, then an edge $s_j\to s_k$ exists with emission $x$ and with transition probability $P(x|j)$. Note that when appending a particular symbol $x$ to any of the pasts $\{\past{x}\}$ in a given causal state $s_j$ their concatenated pasts $\{\past{x}x\}$ will all belong to the same causal state $s_k$. That is, given the initial causal state and output symbol, the subsequent causal state is uniquely determined. This follows directly from the causal equivalence relation, and within computational mechanics is referred to as unifilarity.

The corresponding HMM is referred to as the $\varepsilon$-\emph{machine} of the process, and is the (classically) provably minimal -- in terms of number of states and their information content -- representation of the process. Due to its priveliged position, this information content -- which can be seen as the amount of information that must be tracked about the past in order to produce a statistically-accurate future set of outputs -- is taken in computational mechanics as a measure of structure, called the \emph{statistical complexity} $C_\mu$:
\begin{equation}
\label{Eq.appCmu}
	C_\mu := H[P(s_j)] = - \sum_{s_j \in \mathcal{S}} P(s_j) \log_2 [P(s_j)],
\end{equation}
where $P(s_j)=\sum_{\past{x}\in s_j}P(\past{x})$ is the probability that an observed past belongs to $s_j$. This quantity can also more abstractly be thought of as the information communicated from the past of the process to the future.

There exist a number of methods for reconstructing $\varepsilon$-machines from data~\cite{Crutchfield1989, Shalizi2004, Strelioff2014}; we do not detail these here. Further background and details on computational mechanics can be found within e.g.~\cite{Crutchfield1989, shalizi2001computational, Crutchfield2011, Shalizi2001}.

\begin{figure}[H]
	\centering
	\includegraphics[width=0.5\linewidth]{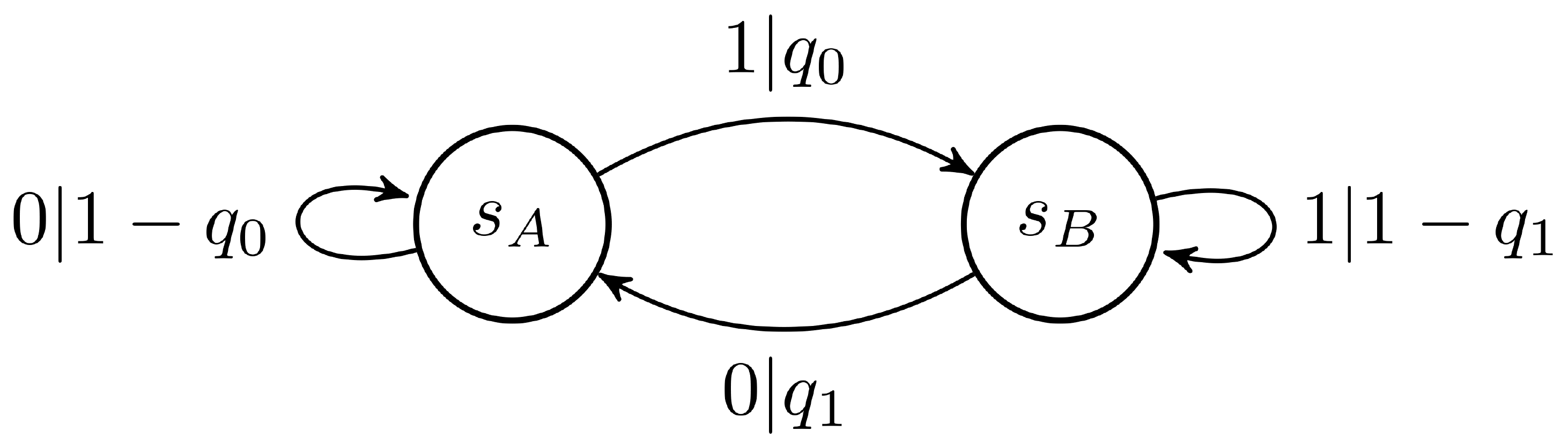}
	\caption{General form of a two-symbol Markov chain. Aside from special parameter sets the states of the chain represent the causal states of the process. The notation $x|p$ on an edge indicates the transition occurs with probability $p$, accompanied by an emission $x$.}
	\label{fig.perturbedcoin}
\end{figure}

As an example of how computational mechanics can be deployed, consider a two-symbol Markov chain (sometimes referred to as the perturbed coin~\cite{Gu2012}), as illustrated in \figref{fig.perturbedcoin}. Given that the process is Markovian, we can immediately conclude that every past for which the most recent symbol is the same can be grouped into the same causal state; in other words, the causal states are either the states of the Markov chain, or a coarse-graining thereof. Indeed, we can see that with the exception of certain special cases, the two states in the chain provide different future statistics, and so form the two causal states of the process.

The aforementioned special cases are when either:
\begin{itemize}
\item $q_0=0$ or $q_1=0$, wherein the process with consist of either an infinite string of 1s, or an infinite string of 0s respectively;
\item $q_0=q_1=1$, wherein the process will be an infinite string of alternating 0s and 1s (i.e., $\ldots010101010101\ldots)$;
\item $q_0=1-q_1$ (and $q_1=1-q_0$), wherein the outputs are stochastic, but both states produce identical future statistics.
\end{itemize}
In all these cases, there is only a single causal state, and trivially $C_\mu=0$.

For all other parameter sets, the steady-state distribution of the causal states can be found from the (normalised) eigenvector of the transition matrix $T$ of the $\varepsilon$-machine with unit eigenvalue. Specifically, we have
\begin{equation}
	T = \begin{pmatrix} 1-q_0 & q_1 \\ q_0 & 1-q_1 \end{pmatrix},
\end{equation}
for which the corresponding eigenvector is $(q_1,q_0)^T/(q_0+q_1)$. The Shannon entropy of the elements of this vector then yield the statistical complexity:
\begin{equation}
C_\mu=\log_2(q_0+q_1)-\frac{q_0}{q_0+q_1}\log_2(q_0)-\frac{q_1}{q_0+q_1}\log_2(q_1).
\end{equation}
Notably, whenever $q_0=q_1\neq\{0,0.5,1\}$ we have that $C_\mu=1$.

\section{Detailed proof of robustness of quantum statistical memory}

Recall from the main text that the Gram matrix $G$ of a quantum state $\rho=\sum_j P_j\ket{\sigma_j}\bra{\sigma_j}$ is defined as $G_{jk}=\sqrt{P_jP_k}\braket{\sigma_j}{\sigma_k}$~\cite{Jozsa2000, Horn2012, Riechers2016}. If one considers a purification of the original state $\ket{\Psi}=\sum_j\sqrt{P_j}\ket{\sigma_j}\ket{j}$, then $\rho$ can be recovered by taking a partial trace over the second subsystem, and $G$ by tracing out the first. The Gram matrix thus has the same spectrum as the original state, and so can be used to calculate functions of this spectrum, such as the entropy. As stated in Eq.~\eqref{eq.gram}, for our construction the Gram matrix is given by
\begin{equation}
	G_{x_{-L:0}x'_{-L:0}} = \sqrt{P(x_{-L:0}) P(x_{-L:0}')} \braket{\varsigma_{x_{-L:0}}}{\varsigma_{x'_{-L:0}}}.
\end{equation}
Using Eq.~\eqref{eq.overlap4}, this can be expanded as
\begin{align}
\label{eq.grammatrixprotocol}
	G_{x_{-L:0}x'_{-L:0}} &=   \sqrt{P(x_{-L:0}) P(x'_{-L:0})}\sum_{x_{0:L}} \sqrt{P(x_{0:L}|x_{-L:0}) P(x_{0:L}|x'_{-L:0})}\nonumber\\
&=\sum_{x_{0:L}}\sqrt{P(x_{-L:0} x_{0:L}) P(x'_{-L:0} x_{0:L})}.
\end{align}

We now examine how this changes when the probabilities $P$ are replaced by their peturbed versions $P^\epsilon:=P+\epsilon\Delta P$. Consider expanding out the square root of the product of two such pertubations:
\begin{align}
\label{eq.p_approx}
	\sqrt{P^\epsilon Q^\epsilon} &= \sqrt{(P + \epsilon  \Delta P)(Q + \epsilon  \Delta Q)}\nonumber \\
	&= \sqrt{PQ} \sqrt{1 + \epsilon\left(\frac{\Delta P}{P} + \frac{\Delta Q}{Q}\right) + \epsilon^2\frac{\Delta P \Delta Q}{PQ} }\nonumber \\
	&\approx \sqrt{PQ} + \epsilon\frac{\sqrt{PQ}}{2} \left( \frac{\Delta P}{P} + \frac{\Delta Q}{Q} \right) + O(\epsilon^2).
\end{align}
Substituting this into Eq.~\eqref{eq.grammatrixprotocol}, we obtain
\begin{align}
\label{eq.grammatrixperturbation1}
	G^\epsilon_{x_{-L:0}x'_{-L:0}} &= \sum_{x_{0:L}} \sqrt{P^\epsilon(x_{-L:0}x_{0:L}) P^\epsilon(x'_{-L:0}x_{0:L})} \nonumber \\
		&\approx\sum_{x_{0:L}}\sqrt{P(x_{-L:0}x_{0:L}) P(x'_{-L:0}x_{0:L})} \nonumber\\
	 &+ \epsilon\sum_{x_{0:L}}\frac{\sqrt{P(x_{-L:0}x_{0:L}) P(x'_{-L:0}x_{0:L})} }{2} \left( \frac{\Delta P(x_{-L:0}x_{0:L})}{P(x_{-L:0}x_{0:L})} + \frac{\Delta P(x'_{-L:0}x_{0:L})}{P((x'_{-L:0}x_{0:L})} \right) \nonumber\\ &+ O(\epsilon^2).
\end{align}
Thus, we can write 
\begin{equation}
\label{eq.gramperturb}
G^\epsilon_{x_{-L:0}x'_{-L:0}}\approx G_{x_{-L:0}x'_{-L:0}} + \epsilon\Delta G_{x_{-L:0}x'_{-L:0}}+O(\epsilon^2),
\end{equation}
where 
\begin{equation}
\label{eq.delG}
\Delta G_{x_{-L:0}x'_{-L:0}} =\sum_{x_{0:L}}\frac{\sqrt{P(x_{-L:0}x_{0:L}) P(x'_{-L:0}x_{0:L})} }{2} \left( \frac{\Delta P(x_{-L:0}x_{0:L})}{P(x_{-L:0}x_{0:L})} + \frac{\Delta P(x'_{-L:0}x_{0:L})}{P((x'_{-L:0}x_{0:L})} \right).
\end{equation}

From Weyl's inequality, it then follows that the perturbation to the eigenvalues of $G$ are bounded by the spectral norm of $\epsilon\Delta G$~\cite{Horn2012}. Clearly, this norm scales with $\epsilon$, and so the perturbation to the spectrum of $G$ varies continuously with $\epsilon$. Finally, since the von Neumann entropy of a quantum state is given by the Shannon entropy of its spectrum -- and is a continuous function of it~\cite{Nielsen2010} -- the quantum statistical memory is smoothly deformed by the pertubation, and so is robust.

\section{Scaling of statistical noise} 

We now examine the effects of statistical noise in our estimates of word probabilities on $\tilde{C}_q$. These fluctuations can be considered as a (stochastic) pertubation, i.e.~$\tilde{P}=P+\epsilon\Delta P$, allowing us to employ the results above. We will set $\epsilon$ to 1, folding the full scaling of the perturbation with $L$ and $N$ into $\Delta P$.

Recall that the corrections to the eigenvalues arising from the perturbation are bounded by the spectral norm of $\Delta G$ -- i.e., its largest eigenvalue -- which in turn is bounded by the product of the dimension of the matrix with its largest element. The elements of this matrix are given in Eq.~\eqref{eq.delG}; to assess how they scale we will replace the statistical variables by their standard errors. Since the word probabilities $P(X_{-L:L})$ are described by binomial distributions (a randomly selected string can be assigned as either being the given word $x_{-L:L}$ or not), the standard error is given by
\begin{equation}
\sigma_{\bar{P}(x_{-L:L})}=\sqrt{\frac{P(x_{-L:L})(1-P(x_{-L:L}))}{N}}.
\end{equation}

Inserting this into Eq.~\eqref{eq.delG} we obtain
\begin{equation}
\Delta G_{x_{-L:0}x'_{-L:0}} =\frac{1}{2\sqrt{N}}\sum_{x_{0:L}}\sqrt{P(x'_{-L:0}x_{0:L})(1-P(x_{-L:0}x_{0:L}))}+\sqrt{P(x_{-L:0}x_{0:L})(1-P(x'_{-L:0}x_{0:L}))}.
\end{equation}
The probability of obtaining a given word of length $L$ falls off approximately exponentially with $L$. Let us assume all long words are roughly evenly distributed, and take $P(X_{-L:L})\sim|\mathcal{A}|^{-2L}$. Considering that there are roughly $|\mathcal{A}|^L$ terms in the sum, we have $\Delta G_{x_{-L:0}x'_{-L:0}} \sim 1/\sqrt{N}$. Finally, since the dimension of the matrix scales as $|\mathcal{A}|^L$, the spectral norm (and thus the bound on the size of perturbations to the spectrum of the Gram matrix) scales $\sim |\mathcal{A}|^{L}/\sqrt{N}$.

\twocolumngrid

\bibliography{ref}

\begin{thebibliography}{45}%
\makeatletter
\providecommand \@ifxundefined [1]{%
 \@ifx{#1\undefined}
}%
\providecommand \@ifnum [1]{%
 \ifnum #1\expandafter \@firstoftwo
 \else \expandafter \@secondoftwo
 \fi
}%
\providecommand \@ifx [1]{%
 \ifx #1\expandafter \@firstoftwo
 \else \expandafter \@secondoftwo
 \fi
}%
\providecommand \natexlab [1]{#1}%
\providecommand \enquote  [1]{``#1''}%
\providecommand \bibnamefont  [1]{#1}%
\providecommand \bibfnamefont [1]{#1}%
\providecommand \citenamefont [1]{#1}%
\providecommand \href@noop [0]{\@secondoftwo}%
\providecommand \href [0]{\begingroup \@sanitize@url \@href}%
\providecommand \@href[1]{\@@startlink{#1}\@@href}%
\providecommand \@@href[1]{\endgroup#1\@@endlink}%
\providecommand \@sanitize@url [0]{\catcode `\\12\catcode `\$12\catcode
  `\&12\catcode `\#12\catcode `\^12\catcode `\_12\catcode `\%12\relax}%
\providecommand \@@startlink[1]{}%
\providecommand \@@endlink[0]{}%
\providecommand \url  [0]{\begingroup\@sanitize@url \@url }%
\providecommand \@url [1]{\endgroup\@href {#1}{\urlprefix }}%
\providecommand \urlprefix  [0]{URL }%
\providecommand \Eprint [0]{\href }%
\providecommand \doibase [0]{http://dx.doi.org/}%
\providecommand \selectlanguage [0]{\@gobble}%
\providecommand \bibinfo  [0]{\@secondoftwo}%
\providecommand \bibfield  [0]{\@secondoftwo}%
\providecommand \translation [1]{[#1]}%
\providecommand \BibitemOpen [0]{}%
\providecommand \bibitemStop [0]{}%
\providecommand \bibitemNoStop [0]{.\EOS\space}%
\providecommand \EOS [0]{\spacefactor3000\relax}%
\providecommand \BibitemShut  [1]{\csname bibitem#1\endcsname}%
\let\auto@bib@innerbib\@empty
\bibitem [{\citenamefont {Lynch}(2008)}]{Lynch2008}%
  \BibitemOpen
  \bibfield  {author} {\bibinfo {author} {\bibfnamefont {P.}~\bibnamefont
  {Lynch}},\ }\href {\doibase 10.1016/j.jcp.2007.02.034} {\bibfield  {journal}
  {\bibinfo  {journal} {Journal of Computational Physics}\ }\textbf {\bibinfo
  {volume} {227}},\ \bibinfo {pages} {3431} (\bibinfo {year}
  {2008})}\BibitemShut {NoStop}%
\bibitem [{\citenamefont {Bauer}\ \emph {et~al.}(2015)\citenamefont {Bauer},
  \citenamefont {Thorpe},\ and\ \citenamefont {Brunet}}]{Bauer2015}%
  \BibitemOpen
  \bibfield  {author} {\bibinfo {author} {\bibfnamefont {P.}~\bibnamefont
  {Bauer}}, \bibinfo {author} {\bibfnamefont {A.}~\bibnamefont {Thorpe}}, \
  and\ \bibinfo {author} {\bibfnamefont {G.}~\bibnamefont {Brunet}},\ }\href
  {https://doi.org/10.1038/nature14956 http://10.0.4.14/nature14956} {\bibfield
   {journal} {\bibinfo  {journal} {Nature}\ }\textbf {\bibinfo {volume}
  {525}},\ \bibinfo {pages} {47} (\bibinfo {year} {2015})}\BibitemShut
  {NoStop}%
\bibitem [{\citenamefont {Pavlos}\ \emph {et~al.}(2015)\citenamefont {Pavlos},
  \citenamefont {Karakatsanis}, \citenamefont {Iliopoulos}, \citenamefont
  {Pavlos}, \citenamefont {Xenakis}, \citenamefont {Clark}, \citenamefont
  {Duke},\ and\ \citenamefont {Monos}}]{Pavlos2015}%
  \BibitemOpen
  \bibfield  {author} {\bibinfo {author} {\bibfnamefont {G.~P.}\ \bibnamefont
  {Pavlos}}, \bibinfo {author} {\bibfnamefont {L.~P.}\ \bibnamefont
  {Karakatsanis}}, \bibinfo {author} {\bibfnamefont {A.~C.}\ \bibnamefont
  {Iliopoulos}}, \bibinfo {author} {\bibfnamefont {E.~G.}\ \bibnamefont
  {Pavlos}}, \bibinfo {author} {\bibfnamefont {M.~N.}\ \bibnamefont {Xenakis}},
  \bibinfo {author} {\bibfnamefont {P.}~\bibnamefont {Clark}}, \bibinfo
  {author} {\bibfnamefont {J.}~\bibnamefont {Duke}}, \ and\ \bibinfo {author}
  {\bibfnamefont {D.~S.}\ \bibnamefont {Monos}},\ }\href {\doibase
  10.1016/j.physa.2015.06.044} {\bibfield  {journal} {\bibinfo  {journal}
  {Physica A}\ }\textbf {\bibinfo {volume} {438}},\ \bibinfo {pages} {188}
  (\bibinfo {year} {2015})}\BibitemShut {NoStop}%
\bibitem [{\citenamefont {Preis}\ \emph {et~al.}(2012)\citenamefont {Preis},
  \citenamefont {Kenett}, \citenamefont {Stanley}, \citenamefont {Helbing},\
  and\ \citenamefont {Ben-Jacob}}]{Preis2012}%
  \BibitemOpen
  \bibfield  {author} {\bibinfo {author} {\bibfnamefont {T.}~\bibnamefont
  {Preis}}, \bibinfo {author} {\bibfnamefont {D.~Y.}\ \bibnamefont {Kenett}},
  \bibinfo {author} {\bibfnamefont {H.~E.}\ \bibnamefont {Stanley}}, \bibinfo
  {author} {\bibfnamefont {D.}~\bibnamefont {Helbing}}, \ and\ \bibinfo
  {author} {\bibfnamefont {E.}~\bibnamefont {Ben-Jacob}},\ }\href {\doibase
  10.1038/srep00752} {\bibfield  {journal} {\bibinfo  {journal} {Scientific
  Reports}\ }\textbf {\bibinfo {volume} {2}},\ \bibinfo {pages} {1} (\bibinfo
  {year} {2012})}\BibitemShut {NoStop}%
\bibitem [{\citenamefont {Kerner}\ and\ \citenamefont
  {Rehborn}(1996)}]{Kerner1996}%
  \BibitemOpen
  \bibfield  {author} {\bibinfo {author} {\bibfnamefont {B.~S.}\ \bibnamefont
  {Kerner}}\ and\ \bibinfo {author} {\bibfnamefont {H.}~\bibnamefont
  {Rehborn}},\ }\href {\doibase 10.1103/PhysRevE.53.R4275} {\bibfield
  {journal} {\bibinfo  {journal} {Physical Review E}\ }\textbf {\bibinfo
  {volume} {53}},\ \bibinfo {pages} {4275} (\bibinfo {year}
  {1996})}\BibitemShut {NoStop}%
\bibitem [{\citenamefont {Crutchfield}\ and\ \citenamefont
  {Young}(1989)}]{Crutchfield1989}%
  \BibitemOpen
  \bibfield  {author} {\bibinfo {author} {\bibfnamefont {J.~P.}\ \bibnamefont
  {Crutchfield}}\ and\ \bibinfo {author} {\bibfnamefont {K.}~\bibnamefont
  {Young}},\ }\href {\doibase 10.1103/PhysRevLett.63.105} {\bibfield  {journal}
  {\bibinfo  {journal} {Physical Review Letters}\ }\textbf {\bibinfo {volume}
  {63}},\ \bibinfo {pages} {105} (\bibinfo {year} {1989})}\BibitemShut
  {NoStop}%
\bibitem [{\citenamefont {Shalizi}\ and\ \citenamefont
  {Crutchfield}(2001)}]{shalizi2001computational}%
  \BibitemOpen
  \bibfield  {author} {\bibinfo {author} {\bibfnamefont {C.~R.}\ \bibnamefont
  {Shalizi}}\ and\ \bibinfo {author} {\bibfnamefont {J.~P.}\ \bibnamefont
  {Crutchfield}},\ }\href@noop {} {\bibfield  {journal} {\bibinfo  {journal}
  {Journal of Statistical Physics}\ }\textbf {\bibinfo {volume} {104}},\
  \bibinfo {pages} {817} (\bibinfo {year} {2001})}\BibitemShut {NoStop}%
\bibitem [{\citenamefont {Crutchfield}(2011)}]{Crutchfield2011}%
  \BibitemOpen
  \bibfield  {author} {\bibinfo {author} {\bibfnamefont {J.~P.}\ \bibnamefont
  {Crutchfield}},\ }\href {\doibase 10.1038/nphys2190} {\bibfield  {journal}
  {\bibinfo  {journal} {Nature Physics}\ }\textbf {\bibinfo {volume} {8}},\
  \bibinfo {pages} {17} (\bibinfo {year} {2011})}\BibitemShut {NoStop}%
\bibitem [{\citenamefont {Hanson}\ and\ \citenamefont
  {Crutchfield}(1997)}]{Hanson1997}%
  \BibitemOpen
  \bibfield  {author} {\bibinfo {author} {\bibfnamefont {J.~E.}\ \bibnamefont
  {Hanson}}\ and\ \bibinfo {author} {\bibfnamefont {J.~P.}\ \bibnamefont
  {Crutchfield}},\ }\href@noop {} {\bibfield  {journal} {\bibinfo  {journal}
  {Physica D: Nonlinear Phenomena}\ }\textbf {\bibinfo {volume} {103}},\
  \bibinfo {pages} {169} (\bibinfo {year} {1997})}\BibitemShut {NoStop}%
\bibitem [{\citenamefont {Gon{\c{c}}alves}\ \emph {et~al.}(1998)\citenamefont
  {Gon{\c{c}}alves}, \citenamefont {Pinto}, \citenamefont {Sartorelli},\ and\
  \citenamefont {{De Oliveira}}}]{Goncalves1998}%
  \BibitemOpen
  \bibfield  {author} {\bibinfo {author} {\bibfnamefont {W.~M.}\ \bibnamefont
  {Gon{\c{c}}alves}}, \bibinfo {author} {\bibfnamefont {R.~D.}\ \bibnamefont
  {Pinto}}, \bibinfo {author} {\bibfnamefont {J.~C.}\ \bibnamefont
  {Sartorelli}}, \ and\ \bibinfo {author} {\bibfnamefont {M.~J.}\ \bibnamefont
  {{De Oliveira}}},\ }\href {\doibase 10.1016/S0378-4371(98)00164-2} {\bibfield
   {journal} {\bibinfo  {journal} {Physica A}\ }\textbf {\bibinfo {volume}
  {257}},\ \bibinfo {pages} {385} (\bibinfo {year} {1998})}\BibitemShut
  {NoStop}%
\bibitem [{\citenamefont {Park}\ \emph {et~al.}(2007)\citenamefont {Park},
  \citenamefont {{Won Lee}}, \citenamefont {Yang}, \citenamefont {Jo},\ and\
  \citenamefont {Moon}}]{Park2007}%
  \BibitemOpen
  \bibfield  {author} {\bibinfo {author} {\bibfnamefont {J.~B.}\ \bibnamefont
  {Park}}, \bibinfo {author} {\bibfnamefont {J.}~\bibnamefont {{Won Lee}}},
  \bibinfo {author} {\bibfnamefont {J.~S.}\ \bibnamefont {Yang}}, \bibinfo
  {author} {\bibfnamefont {H.~H.}\ \bibnamefont {Jo}}, \ and\ \bibinfo {author}
  {\bibfnamefont {H.~T.}\ \bibnamefont {Moon}},\ }\href {\doibase
  10.1016/j.physa.2006.12.042} {\bibfield  {journal} {\bibinfo  {journal}
  {Physica A}\ }\textbf {\bibinfo {volume} {379}},\ \bibinfo {pages} {179}
  (\bibinfo {year} {2007})}\BibitemShut {NoStop}%
\bibitem [{\citenamefont {Haslinger}\ \emph {et~al.}(2010)\citenamefont
  {Haslinger}, \citenamefont {Klinkner},\ and\ \citenamefont
  {Shalizi}}]{Haslinger2010}%
  \BibitemOpen
  \bibfield  {author} {\bibinfo {author} {\bibfnamefont {R.}~\bibnamefont
  {Haslinger}}, \bibinfo {author} {\bibfnamefont {K.~L.}\ \bibnamefont
  {Klinkner}}, \ and\ \bibinfo {author} {\bibfnamefont {C.~R.}\ \bibnamefont
  {Shalizi}},\ }\href {\doibase 10.1162/neco.2009.12-07-678} {\bibfield
  {journal} {\bibinfo  {journal} {Neural Computation}\ }\textbf {\bibinfo
  {volume} {22}},\ \bibinfo {pages} {121} (\bibinfo {year} {2010})}\BibitemShut
  {NoStop}%
\bibitem [{\citenamefont {Montanaro}(2016)}]{montanaro2016quantum}%
  \BibitemOpen
  \bibfield  {author} {\bibinfo {author} {\bibfnamefont {A.}~\bibnamefont
  {Montanaro}},\ }\href@noop {} {\bibfield  {journal} {\bibinfo  {journal} {npj
  Quantum Information}\ }\textbf {\bibinfo {volume} {2}},\ \bibinfo {pages}
  {15023} (\bibinfo {year} {2016})}\BibitemShut {NoStop}%
\bibitem [{\citenamefont {Bennett}\ and\ \citenamefont
  {Brassard}(2014)}]{bennett2014quantum}%
  \BibitemOpen
  \bibfield  {author} {\bibinfo {author} {\bibfnamefont {C.~H.}\ \bibnamefont
  {Bennett}}\ and\ \bibinfo {author} {\bibfnamefont {G.}~\bibnamefont
  {Brassard}},\ }\href@noop {} {\bibfield  {journal} {\bibinfo  {journal}
  {Theoretical Computer Science}\ }\textbf {\bibinfo {volume} {560}},\ \bibinfo
  {pages} {7} (\bibinfo {year} {2014})}\BibitemShut {NoStop}%
\bibitem [{\citenamefont {Gu}\ \emph {et~al.}(2012)\citenamefont {Gu},
  \citenamefont {Wiesner}, \citenamefont {Rieper},\ and\ \citenamefont
  {Vedral}}]{Gu2012}%
  \BibitemOpen
  \bibfield  {author} {\bibinfo {author} {\bibfnamefont {M.}~\bibnamefont
  {Gu}}, \bibinfo {author} {\bibfnamefont {K.}~\bibnamefont {Wiesner}},
  \bibinfo {author} {\bibfnamefont {E.}~\bibnamefont {Rieper}}, \ and\ \bibinfo
  {author} {\bibfnamefont {V.}~\bibnamefont {Vedral}},\ }\href {\doibase
  10.1038/ncomms1761} {\bibfield  {journal} {\bibinfo  {journal} {Nature
  Communications}\ }\textbf {\bibinfo {volume} {3}},\ \bibinfo {pages} {762}
  (\bibinfo {year} {2012})}\BibitemShut {NoStop}%
\bibitem [{\citenamefont {Mahoney}\ \emph {et~al.}(2016)\citenamefont
  {Mahoney}, \citenamefont {Aghamohammadi},\ and\ \citenamefont
  {Crutchfield}}]{Mahoney2016}%
  \BibitemOpen
  \bibfield  {author} {\bibinfo {author} {\bibfnamefont {J.~R.}\ \bibnamefont
  {Mahoney}}, \bibinfo {author} {\bibfnamefont {C.}~\bibnamefont
  {Aghamohammadi}}, \ and\ \bibinfo {author} {\bibfnamefont {J.~P.}\
  \bibnamefont {Crutchfield}},\ }\href {\doibase 10.1038/srep20495} {\bibfield
  {journal} {\bibinfo  {journal} {Scientific Reports}\ }\textbf {\bibinfo
  {volume} {6}},\ \bibinfo {pages} {20495} (\bibinfo {year}
  {2016})}\BibitemShut {NoStop}%
\bibitem [{\citenamefont {Riechers}\ \emph {et~al.}(2016)\citenamefont
  {Riechers}, \citenamefont {Mahoney}, \citenamefont {Aghamohammadi},\ and\
  \citenamefont {Crutchfield}}]{Riechers2016}%
  \BibitemOpen
  \bibfield  {author} {\bibinfo {author} {\bibfnamefont {P.~M.}\ \bibnamefont
  {Riechers}}, \bibinfo {author} {\bibfnamefont {J.~R.}\ \bibnamefont
  {Mahoney}}, \bibinfo {author} {\bibfnamefont {C.}~\bibnamefont
  {Aghamohammadi}}, \ and\ \bibinfo {author} {\bibfnamefont {J.~P.}\
  \bibnamefont {Crutchfield}},\ }\href {\doibase 10.1103/PhysRevA.93.052317}
  {\bibfield  {journal} {\bibinfo  {journal} {Physical Review A}\ }\textbf
  {\bibinfo {volume} {93}},\ \bibinfo {pages} {052317} (\bibinfo {year}
  {2016})}\BibitemShut {NoStop}%
\bibitem [{\citenamefont {Thompson}\ \emph {et~al.}(2017)\citenamefont
  {Thompson}, \citenamefont {Garner}, \citenamefont {Vedral},\ and\
  \citenamefont {Gu}}]{Thompson2017}%
  \BibitemOpen
  \bibfield  {author} {\bibinfo {author} {\bibfnamefont {J.}~\bibnamefont
  {Thompson}}, \bibinfo {author} {\bibfnamefont {A.~J.~P.}\ \bibnamefont
  {Garner}}, \bibinfo {author} {\bibfnamefont {V.}~\bibnamefont {Vedral}}, \
  and\ \bibinfo {author} {\bibfnamefont {M.}~\bibnamefont {Gu}},\ }\href
  {\doibase 10.1038/s41534-016-0001-3} {\bibfield  {journal} {\bibinfo
  {journal} {npj Quantum Information}\ }\textbf {\bibinfo {volume} {3}},\
  \bibinfo {pages} {6} (\bibinfo {year} {2017})}\BibitemShut {NoStop}%
\bibitem [{\citenamefont {Elliott}\ and\ \citenamefont
  {Gu}(2018)}]{Elliott2018a}%
  \BibitemOpen
  \bibfield  {author} {\bibinfo {author} {\bibfnamefont {T.~J.}\ \bibnamefont
  {Elliott}}\ and\ \bibinfo {author} {\bibfnamefont {M.}~\bibnamefont {Gu}},\
  }\href {\doibase 10.1038/s41534-018-0064-4} {\bibfield  {journal} {\bibinfo
  {journal} {npj Quantum Information}\ }\textbf {\bibinfo {volume} {4}},\
  \bibinfo {pages} {18} (\bibinfo {year} {2018})}\BibitemShut {NoStop}%
\bibitem [{\citenamefont {Binder}\ \emph {et~al.}(2018)\citenamefont {Binder},
  \citenamefont {Thompson},\ and\ \citenamefont {Gu}}]{Binder2018}%
  \BibitemOpen
  \bibfield  {author} {\bibinfo {author} {\bibfnamefont {F.~C.}\ \bibnamefont
  {Binder}}, \bibinfo {author} {\bibfnamefont {J.}~\bibnamefont {Thompson}}, \
  and\ \bibinfo {author} {\bibfnamefont {M.}~\bibnamefont {Gu}},\ }\href
  {\doibase 10.1103/PhysRevLett.120.240502} {\bibfield  {journal} {\bibinfo
  {journal} {Physical Review Letters}\ }\textbf {\bibinfo {volume} {120}},\
  \bibinfo {pages} {240502} (\bibinfo {year} {2018})},\ \Eprint
  {http://arxiv.org/abs/1709.02375} {arXiv:1709.02375} \BibitemShut {NoStop}%
\bibitem [{\citenamefont {Elliott}\ \emph
  {et~al.}(2019{\natexlab{a}})\citenamefont {Elliott}, \citenamefont {Garner},\
  and\ \citenamefont {Gu}}]{Elliott2018b}%
  \BibitemOpen
  \bibfield  {author} {\bibinfo {author} {\bibfnamefont {T.~J.}\ \bibnamefont
  {Elliott}}, \bibinfo {author} {\bibfnamefont {A.~J.~P.}\ \bibnamefont
  {Garner}}, \ and\ \bibinfo {author} {\bibfnamefont {M.}~\bibnamefont {Gu}},\
  }\href {\doibase 10.1088/1367-2630/aaf824} {\bibfield  {journal} {\bibinfo
  {journal} {New Journal of Physics}\ }\textbf {\bibinfo {volume} {21}},\
  \bibinfo {pages} {013021} (\bibinfo {year} {2019}{\natexlab{a}})}\BibitemShut
  {NoStop}%
\bibitem [{\citenamefont {Liu}\ \emph {et~al.}(2019)\citenamefont {Liu},
  \citenamefont {Elliott}, \citenamefont {Binder}, \citenamefont {{Di
  Franco}},\ and\ \citenamefont {Gu}}]{Liu2019}%
  \BibitemOpen
  \bibfield  {author} {\bibinfo {author} {\bibfnamefont {Q.}~\bibnamefont
  {Liu}}, \bibinfo {author} {\bibfnamefont {T.~J.}\ \bibnamefont {Elliott}},
  \bibinfo {author} {\bibfnamefont {F.~C.}\ \bibnamefont {Binder}}, \bibinfo
  {author} {\bibfnamefont {C.}~\bibnamefont {{Di Franco}}}, \ and\ \bibinfo
  {author} {\bibfnamefont {M.}~\bibnamefont {Gu}},\ }\href {\doibase
  10.1103/PhysRevA.99.062110} {\bibfield  {journal} {\bibinfo  {journal}
  {Physical Review A}\ }\textbf {\bibinfo {volume} {99}},\ \bibinfo {pages}
  {062110} (\bibinfo {year} {2019})}\BibitemShut {NoStop}%
\bibitem [{\citenamefont {Garner}\ \emph {et~al.}(2017)\citenamefont {Garner},
  \citenamefont {Liu}, \citenamefont {Thompson},\ and\ \citenamefont
  {Vedral}}]{Garner2017a}%
  \BibitemOpen
  \bibfield  {author} {\bibinfo {author} {\bibfnamefont {A.~J.~P.}\
  \bibnamefont {Garner}}, \bibinfo {author} {\bibfnamefont {Q.}~\bibnamefont
  {Liu}}, \bibinfo {author} {\bibfnamefont {J.}~\bibnamefont {Thompson}}, \
  and\ \bibinfo {author} {\bibfnamefont {V.}~\bibnamefont {Vedral}},\
  }\href@noop {} {\bibfield  {journal} {\bibinfo  {journal} {New Journal of
  Physics}\ }\textbf {\bibinfo {volume} {19}},\ \bibinfo {pages} {103009}
  (\bibinfo {year} {2017})}\BibitemShut {NoStop}%
\bibitem [{\citenamefont {Aghamohammadi}\ \emph
  {et~al.}(2017{\natexlab{a}})\citenamefont {Aghamohammadi}, \citenamefont
  {Mahoney},\ and\ \citenamefont {Crutchfield}}]{Aghamohammadi2017}%
  \BibitemOpen
  \bibfield  {author} {\bibinfo {author} {\bibfnamefont {C.}~\bibnamefont
  {Aghamohammadi}}, \bibinfo {author} {\bibfnamefont {J.~R.}\ \bibnamefont
  {Mahoney}}, \ and\ \bibinfo {author} {\bibfnamefont {J.~P.}\ \bibnamefont
  {Crutchfield}},\ }\href {\doibase 10.1038/s41598-017-04928-7} {\bibfield
  {journal} {\bibinfo  {journal} {Scientific Reports}\ }\textbf {\bibinfo
  {volume} {7}},\ \bibinfo {pages} {6735} (\bibinfo {year}
  {2017}{\natexlab{a}})}\BibitemShut {NoStop}%
\bibitem [{\citenamefont {Thompson}\ \emph {et~al.}(2018)\citenamefont
  {Thompson}, \citenamefont {Garner}, \citenamefont {Mahoney}, \citenamefont
  {Crutchfield}, \citenamefont {Vedral},\ and\ \citenamefont
  {Gu}}]{Thompson2018a}%
  \BibitemOpen
  \bibfield  {author} {\bibinfo {author} {\bibfnamefont {J.}~\bibnamefont
  {Thompson}}, \bibinfo {author} {\bibfnamefont {A.~J.}\ \bibnamefont
  {Garner}}, \bibinfo {author} {\bibfnamefont {J.~R.}\ \bibnamefont {Mahoney}},
  \bibinfo {author} {\bibfnamefont {J.~P.}\ \bibnamefont {Crutchfield}},
  \bibinfo {author} {\bibfnamefont {V.}~\bibnamefont {Vedral}}, \ and\ \bibinfo
  {author} {\bibfnamefont {M.}~\bibnamefont {Gu}},\ }\href {\doibase
  10.1103/PhysRevX.8.031013} {\bibfield  {journal} {\bibinfo  {journal}
  {Physical Review X}\ }\textbf {\bibinfo {volume} {8}},\ \bibinfo {pages}
  {031013} (\bibinfo {year} {2018})}\BibitemShut {NoStop}%
\bibitem [{\citenamefont {Elliott}\ \emph
  {et~al.}(2019{\natexlab{b}})\citenamefont {Elliott}, \citenamefont {Yang},
  \citenamefont {Binder}, \citenamefont {Garner}, \citenamefont {Thompson},\
  and\ \citenamefont {Gu}}]{elliott2019extreme}%
  \BibitemOpen
  \bibfield  {author} {\bibinfo {author} {\bibfnamefont {T.~J.}\ \bibnamefont
  {Elliott}}, \bibinfo {author} {\bibfnamefont {C.}~\bibnamefont {Yang}},
  \bibinfo {author} {\bibfnamefont {F.~C.}\ \bibnamefont {Binder}}, \bibinfo
  {author} {\bibfnamefont {A.~J.~P.}\ \bibnamefont {Garner}}, \bibinfo {author}
  {\bibfnamefont {J.}~\bibnamefont {Thompson}}, \ and\ \bibinfo {author}
  {\bibfnamefont {M.}~\bibnamefont {Gu}},\ }\href@noop {} {\bibfield  {journal}
  {\bibinfo  {journal} {arXiv preprint arXiv:1909.02817}\ } (\bibinfo {year}
  {2019}{\natexlab{b}})}\BibitemShut {NoStop}%
\bibitem [{\citenamefont {Palsson}\ \emph {et~al.}(2017)\citenamefont
  {Palsson}, \citenamefont {Gu}, \citenamefont {Ho}, \citenamefont {Wiseman},\
  and\ \citenamefont {Pryde}}]{Palsson2017a}%
  \BibitemOpen
  \bibfield  {author} {\bibinfo {author} {\bibfnamefont {M.~S.}\ \bibnamefont
  {Palsson}}, \bibinfo {author} {\bibfnamefont {M.}~\bibnamefont {Gu}},
  \bibinfo {author} {\bibfnamefont {J.}~\bibnamefont {Ho}}, \bibinfo {author}
  {\bibfnamefont {H.~M.}\ \bibnamefont {Wiseman}}, \ and\ \bibinfo {author}
  {\bibfnamefont {G.~J.}\ \bibnamefont {Pryde}},\ }\href {\doibase
  10.1126/sciadv.1601302} {\bibfield  {journal} {\bibinfo  {journal} {Science
  Advances}\ }\textbf {\bibinfo {volume} {3}} (\bibinfo {year} {2017}),\
  10.1126/sciadv.1601302}\BibitemShut {NoStop}%
\bibitem [{\citenamefont {Jouneghani}\ \emph {et~al.}(2017)\citenamefont
  {Jouneghani}, \citenamefont {Gu}, \citenamefont {Ho}, \citenamefont
  {Thompson}, \citenamefont {Suen}, \citenamefont {Wiseman},\ and\
  \citenamefont {Pryde}}]{Jouneghani2017}%
  \BibitemOpen
  \bibfield  {author} {\bibinfo {author} {\bibfnamefont {F.~G.}\ \bibnamefont
  {Jouneghani}}, \bibinfo {author} {\bibfnamefont {M.}~\bibnamefont {Gu}},
  \bibinfo {author} {\bibfnamefont {J.}~\bibnamefont {Ho}}, \bibinfo {author}
  {\bibfnamefont {J.}~\bibnamefont {Thompson}}, \bibinfo {author}
  {\bibfnamefont {W.~Y.}\ \bibnamefont {Suen}}, \bibinfo {author}
  {\bibfnamefont {H.~M.}\ \bibnamefont {Wiseman}}, \ and\ \bibinfo {author}
  {\bibfnamefont {G.~J.}\ \bibnamefont {Pryde}},\ }\href
  {http://arxiv.org/abs/1711.03661} {\  (\bibinfo {year} {2017})},\ \Eprint
  {http://arxiv.org/abs/1711.03661} {arXiv:1711.03661} \BibitemShut {NoStop}%
\bibitem [{\citenamefont {Ghafari}\ \emph {et~al.}(2019)\citenamefont
  {Ghafari}, \citenamefont {Tischler}, \citenamefont {Thompson}, \citenamefont
  {Gu}, \citenamefont {Shalm}, \citenamefont {Verma}, \citenamefont {Nam},
  \citenamefont {Patel}, \citenamefont {Wiseman},\ and\ \citenamefont
  {Pryde}}]{Ghafari2018}%
  \BibitemOpen
  \bibfield  {author} {\bibinfo {author} {\bibfnamefont {F.}~\bibnamefont
  {Ghafari}}, \bibinfo {author} {\bibfnamefont {N.}~\bibnamefont {Tischler}},
  \bibinfo {author} {\bibfnamefont {J.}~\bibnamefont {Thompson}}, \bibinfo
  {author} {\bibfnamefont {M.}~\bibnamefont {Gu}}, \bibinfo {author}
  {\bibfnamefont {L.~K.}\ \bibnamefont {Shalm}}, \bibinfo {author}
  {\bibfnamefont {V.~B.}\ \bibnamefont {Verma}}, \bibinfo {author}
  {\bibfnamefont {S.~W.}\ \bibnamefont {Nam}}, \bibinfo {author} {\bibfnamefont
  {R.~B.}\ \bibnamefont {Patel}}, \bibinfo {author} {\bibfnamefont {H.~M.}\
  \bibnamefont {Wiseman}}, \ and\ \bibinfo {author} {\bibfnamefont {G.~J.}\
  \bibnamefont {Pryde}},\ }\href@noop {} {\bibfield  {journal} {\bibinfo
  {journal} {Physical Review X}\ }\textbf {\bibinfo {volume} {9}},\ \bibinfo
  {pages} {041013} (\bibinfo {year} {2019})}\BibitemShut {NoStop}%
\bibitem [{\citenamefont {Tan}\ \emph {et~al.}(2014)\citenamefont {Tan},
  \citenamefont {{R. Terno}}, \citenamefont {Thompson}, \citenamefont
  {Vedral},\ and\ \citenamefont {Gu}}]{Tan2014}%
  \BibitemOpen
  \bibfield  {author} {\bibinfo {author} {\bibfnamefont {R.}~\bibnamefont
  {Tan}}, \bibinfo {author} {\bibfnamefont {D.}~\bibnamefont {{R. Terno}}},
  \bibinfo {author} {\bibfnamefont {J.}~\bibnamefont {Thompson}}, \bibinfo
  {author} {\bibfnamefont {V.}~\bibnamefont {Vedral}}, \ and\ \bibinfo {author}
  {\bibfnamefont {M.}~\bibnamefont {Gu}},\ }\href {\doibase
  10.1140/epjp/i2014-14191-2} {\bibfield  {journal} {\bibinfo  {journal}
  {European Physical Journal Plus}\ }\textbf {\bibinfo {volume} {129:191}}
  (\bibinfo {year} {2014}),\ 10.1140/epjp/i2014-14191-2}\BibitemShut {NoStop}%
\bibitem [{\citenamefont {Suen}\ \emph {et~al.}(2017)\citenamefont {Suen},
  \citenamefont {Thompson}, \citenamefont {Garner}, \citenamefont {Vedral},\
  and\ \citenamefont {Gu}}]{Suen2017}%
  \BibitemOpen
  \bibfield  {author} {\bibinfo {author} {\bibfnamefont {W.~Y.}\ \bibnamefont
  {Suen}}, \bibinfo {author} {\bibfnamefont {J.}~\bibnamefont {Thompson}},
  \bibinfo {author} {\bibfnamefont {A.~J.~P.}\ \bibnamefont {Garner}}, \bibinfo
  {author} {\bibfnamefont {V.}~\bibnamefont {Vedral}}, \ and\ \bibinfo {author}
  {\bibfnamefont {M.}~\bibnamefont {Gu}},\ }\href
  {https://doi.org/10.22331/q-2017-08-11-25} {\bibfield  {journal} {\bibinfo
  {journal} {Quantum}\ }\textbf {\bibinfo {volume} {1}},\ \bibinfo {pages} {25}
  (\bibinfo {year} {2017})}\BibitemShut {NoStop}%
\bibitem [{\citenamefont {Aghamohammadi}\ \emph
  {et~al.}(2017{\natexlab{b}})\citenamefont {Aghamohammadi}, \citenamefont
  {Mahoney},\ and\ \citenamefont {Crutchfield}}]{Aghamohammadi2017a}%
  \BibitemOpen
  \bibfield  {author} {\bibinfo {author} {\bibfnamefont {C.}~\bibnamefont
  {Aghamohammadi}}, \bibinfo {author} {\bibfnamefont {J.~R.}\ \bibnamefont
  {Mahoney}}, \ and\ \bibinfo {author} {\bibfnamefont {J.~P.}\ \bibnamefont
  {Crutchfield}},\ }\href {\doibase 10.1016/j.physleta.2016.12.036} {\bibfield
  {journal} {\bibinfo  {journal} {Physics Letters A}\ }\textbf {\bibinfo
  {volume} {381}},\ \bibinfo {pages} {1223} (\bibinfo {year}
  {2017}{\natexlab{b}})}\BibitemShut {NoStop}%
\bibitem [{\citenamefont {Suen}\ \emph {et~al.}(2018)\citenamefont {Suen},
  \citenamefont {Elliott}, \citenamefont {Thompson}, \citenamefont {Garner},
  \citenamefont {Mahoney}, \citenamefont {Vedral},\ and\ \citenamefont
  {Gu}}]{Suen2018}%
  \BibitemOpen
  \bibfield  {author} {\bibinfo {author} {\bibfnamefont {W.~Y.}\ \bibnamefont
  {Suen}}, \bibinfo {author} {\bibfnamefont {T.~J.}\ \bibnamefont {Elliott}},
  \bibinfo {author} {\bibfnamefont {J.}~\bibnamefont {Thompson}}, \bibinfo
  {author} {\bibfnamefont {A.~J.~P.}\ \bibnamefont {Garner}}, \bibinfo {author}
  {\bibfnamefont {J.~R.}\ \bibnamefont {Mahoney}}, \bibinfo {author}
  {\bibfnamefont {V.}~\bibnamefont {Vedral}}, \ and\ \bibinfo {author}
  {\bibfnamefont {M.}~\bibnamefont {Gu}},\ }\href
  {http://arxiv.org/abs/1812.09738} {\  (\bibinfo {year} {2018})},\ \Eprint
  {http://arxiv.org/abs/1812.09738} {arXiv:1812.09738} \BibitemShut {NoStop}%
\bibitem [{\citenamefont {Shalizi}\ and\ \citenamefont
  {Klinker}(2004)}]{Shalizi2004}%
  \BibitemOpen
  \bibfield  {author} {\bibinfo {author} {\bibfnamefont {C.~R.}\ \bibnamefont
  {Shalizi}}\ and\ \bibinfo {author} {\bibfnamefont {K.~L.}\ \bibnamefont
  {Klinker}},\ }in\ \href@noop {} {\emph {\bibinfo {booktitle} {Uncertainty in
  Artificial Intelligence: Proceedings of the Twentieth Conference (UAI
  2004)}}},\ \bibinfo {editor} {edited by\ \bibinfo {editor} {\bibnamefont
  {{Max Chickering}}}\ and\ \bibinfo {editor} {\bibfnamefont {J.~Y.}\
  \bibnamefont {Halpern}}}\ (\bibinfo  {publisher} {AUAI Press},\ \bibinfo
  {address} {Arlington, Virginia},\ \bibinfo {year} {2004})\ pp.\ \bibinfo
  {pages} {504--511}\BibitemShut {NoStop}%
\bibitem [{\citenamefont {Strelioff}\ and\ \citenamefont
  {Crutchfield}(2014)}]{Strelioff2014}%
  \BibitemOpen
  \bibfield  {author} {\bibinfo {author} {\bibfnamefont {C.~C.}\ \bibnamefont
  {Strelioff}}\ and\ \bibinfo {author} {\bibfnamefont {J.~P.}\ \bibnamefont
  {Crutchfield}},\ }\href {\doibase 10.1103/PhysRevE.89.042119} {\bibfield
  {journal} {\bibinfo  {journal} {Physical Review E}\ }\textbf {\bibinfo
  {volume} {89}},\ \bibinfo {pages} {042119} (\bibinfo {year}
  {2014})}\BibitemShut {NoStop}%
\bibitem [{\citenamefont {Racca}\ \emph {et~al.}(2007)\citenamefont {Racca},
  \citenamefont {Laio}, \citenamefont {Poggi},\ and\ \citenamefont
  {Ridolfi}}]{Racca2007}%
  \BibitemOpen
  \bibfield  {author} {\bibinfo {author} {\bibfnamefont {E.}~\bibnamefont
  {Racca}}, \bibinfo {author} {\bibfnamefont {F.}~\bibnamefont {Laio}},
  \bibinfo {author} {\bibfnamefont {D.}~\bibnamefont {Poggi}}, \ and\ \bibinfo
  {author} {\bibfnamefont {L.}~\bibnamefont {Ridolfi}},\ }\href {\doibase
  10.1103/PhysRevE.75.011126} {\bibfield  {journal} {\bibinfo  {journal}
  {Physical Review E}\ }\textbf {\bibinfo {volume} {75}},\ \bibinfo {pages}
  {011126} (\bibinfo {year} {2007})}\BibitemShut {NoStop}%
\bibitem [{\citenamefont {Mahoney}\ \emph {et~al.}(2011)\citenamefont
  {Mahoney}, \citenamefont {Ellison}, \citenamefont {James},\ and\
  \citenamefont {Crutchfield}}]{Mahoney2011a}%
  \BibitemOpen
  \bibfield  {author} {\bibinfo {author} {\bibfnamefont {J.~R.}\ \bibnamefont
  {Mahoney}}, \bibinfo {author} {\bibfnamefont {C.~J.}\ \bibnamefont
  {Ellison}}, \bibinfo {author} {\bibfnamefont {R.~G.}\ \bibnamefont {James}},
  \ and\ \bibinfo {author} {\bibfnamefont {J.~P.}\ \bibnamefont
  {Crutchfield}},\ }\href {\doibase 10.1063/1.3637502} {\bibfield  {journal}
  {\bibinfo  {journal} {Chaos}\ }\textbf {\bibinfo {volume} {21}},\ \bibinfo
  {pages} {037112} (\bibinfo {year} {2011})}\BibitemShut {NoStop}%
\bibitem [{\citenamefont {Jozsa}(2000)}]{Jozsa2000}%
  \BibitemOpen
  \bibfield  {author} {\bibinfo {author} {\bibfnamefont {R.}~\bibnamefont
  {Jozsa}},\ }\href@noop {} {\bibfield  {journal} {\bibinfo  {journal}
  {Physical Review A}\ }\textbf {\bibinfo {volume} {62}},\ \bibinfo {pages}
  {012301} (\bibinfo {year} {2000})}\BibitemShut {NoStop}%
\bibitem [{\citenamefont {Horn}\ and\ \citenamefont
  {Johnson}(2012)}]{Horn2012}%
  \BibitemOpen
  \bibfield  {author} {\bibinfo {author} {\bibfnamefont {R.~A.}\ \bibnamefont
  {Horn}}\ and\ \bibinfo {author} {\bibfnamefont {C.~R.}\ \bibnamefont
  {Johnson}},\ }\href {\doibase
  10.1002/1521-3773(20010316)40:6<9823::AID-ANIE9823>3.3.CO;2-C} {\emph
  {\bibinfo {title} {{Matrix Analysis}}}},\ \bibinfo {edition} {2nd}\ ed.\
  (\bibinfo  {publisher} {Cambridge University Press},\ \bibinfo {year}
  {2012})\BibitemShut {NoStop}%
\bibitem [{\citenamefont {Nielsen}\ and\ \citenamefont
  {Chuang}(2010)}]{Nielsen2010}%
  \BibitemOpen
  \bibfield  {author} {\bibinfo {author} {\bibfnamefont {M.~A.}\ \bibnamefont
  {Nielsen}}\ and\ \bibinfo {author} {\bibfnamefont {I.~L.}\ \bibnamefont
  {Chuang}},\ }\href {\doibase 10.1017/CBO9781107415324.004} {\emph {\bibinfo
  {title} {Cambridge}}}\ (\bibinfo {year} {2010})\BibitemShut {NoStop}%
\bibitem [{\citenamefont {Crutchfield}\ \emph {et~al.}(2009)\citenamefont
  {Crutchfield}, \citenamefont {Ellison},\ and\ \citenamefont
  {Mahoney}}]{Crutchfield2009a}%
  \BibitemOpen
  \bibfield  {author} {\bibinfo {author} {\bibfnamefont {J.~P.}\ \bibnamefont
  {Crutchfield}}, \bibinfo {author} {\bibfnamefont {C.~J.}\ \bibnamefont
  {Ellison}}, \ and\ \bibinfo {author} {\bibfnamefont {J.~R.}\ \bibnamefont
  {Mahoney}},\ }\href {\doibase 10.1103/PhysRevLett.103.094101} {\bibfield
  {journal} {\bibinfo  {journal} {Physical Review Letters}\ }\textbf {\bibinfo
  {volume} {103}},\ \bibinfo {pages} {94101} (\bibinfo {year}
  {2009})}\BibitemShut {NoStop}%
\bibitem [{\citenamefont {Mahoney}\ \emph {et~al.}(2009)\citenamefont
  {Mahoney}, \citenamefont {Ellison},\ and\ \citenamefont
  {Crutchfield}}]{Mahoney2009}%
  \BibitemOpen
  \bibfield  {author} {\bibinfo {author} {\bibfnamefont {J.~R.}\ \bibnamefont
  {Mahoney}}, \bibinfo {author} {\bibfnamefont {C.~J.}\ \bibnamefont
  {Ellison}}, \ and\ \bibinfo {author} {\bibfnamefont {J.~P.}\ \bibnamefont
  {Crutchfield}},\ }\href {\doibase 10.1088/1751-8113/42/36/362002} {\bibfield
  {journal} {\bibinfo  {journal} {Journal of Physics A: Mathematical and
  Theoretical}\ }\textbf {\bibinfo {volume} {42}},\ \bibinfo {pages} {362002}
  (\bibinfo {year} {2009})}\BibitemShut {NoStop}%
\bibitem [{\citenamefont {Loomis}\ and\ \citenamefont
  {Crutchfield}(2019)}]{loomis2019strong}%
  \BibitemOpen
  \bibfield  {author} {\bibinfo {author} {\bibfnamefont {S.~P.}\ \bibnamefont
  {Loomis}}\ and\ \bibinfo {author} {\bibfnamefont {J.~P.}\ \bibnamefont
  {Crutchfield}},\ }\href@noop {} {\bibfield  {journal} {\bibinfo  {journal}
  {Journal of Statistical Physics}\ }\textbf {\bibinfo {volume} {176}},\
  \bibinfo {pages} {1317} (\bibinfo {year} {2019})}\BibitemShut {NoStop}%
\bibitem [{\citenamefont {Yang}\ \emph {et~al.}(2019)\citenamefont {Yang},
  \citenamefont {Binder}, \citenamefont {Gu},\ and\ \citenamefont
  {Elliott}}]{yang2019measures}%
  \BibitemOpen
  \bibfield  {author} {\bibinfo {author} {\bibfnamefont {C.}~\bibnamefont
  {Yang}}, \bibinfo {author} {\bibfnamefont {F.~C.}\ \bibnamefont {Binder}},
  \bibinfo {author} {\bibfnamefont {M.}~\bibnamefont {Gu}}, \ and\ \bibinfo
  {author} {\bibfnamefont {T.~J.}\ \bibnamefont {Elliott}},\ }\href@noop {}
  {\bibfield  {journal} {\bibinfo  {journal} {arXiv preprint arXiv:1909.08366}\
  } (\bibinfo {year} {2019})}\BibitemShut {NoStop}%
\bibitem [{\citenamefont {Shalizi}(2001)}]{Shalizi2001}%
  \BibitemOpen
  \bibfield  {author} {\bibinfo {author} {\bibfnamefont {C.~R.}\ \bibnamefont
  {Shalizi}},\ }\href {http://bactra.org/thesis/} {\bibfield  {journal}
  {\bibinfo  {journal} {PhD Thesis}\ } (\bibinfo {year} {2001})}\BibitemShut
  {NoStop}%
\end{thebibliography}%

\end{document}